\documentclass[11pt,letterpaper]{article}

\usepackage[margin=1in]{geometry}
\usepackage[T1]{fontenc}
\usepackage[utf8]{inputenc}
\usepackage{amsmath,amssymb,amsfonts,bm}
\usepackage{graphicx}
\usepackage{subcaption}
\usepackage{tabularx}
\usepackage{array}
\usepackage{multirow}
\usepackage{booktabs}
\usepackage{float}
\usepackage{flafter} 
\usepackage[section]{placeins} 
\usepackage{microtype} 
\usepackage[authoryear,round]{natbib}
\usepackage[hidelinks]{hyperref} 

\newcommand{\includegraphicssafe}[2][]{%
  \IfFileExists{#2}{\includegraphics[#1]{#2}}{%
    \fbox{\parbox[c][0.20\textheight][c]{0.95\linewidth}{\centering \textit{Missing figure file:} \texttt{\detokenize{#2}}}}}%
}

\usepackage{caption}      
\usepackage{xcolor}       
\DeclareMathOperator*{\argmin}{arg\,min}

\newcommand{\vect}[1]{\bm{#1}} 
\newcommand{\mat}[1]{\bm{#1}}

\renewcommand{\arraystretch}{1.15}

\newcommand{\keywords}[1]{%
  \vspace{0.5em}
  \noindent\textbf{Keywords:} #1
}

\title{CredibleDFGO: Differentiable Factor Graph Optimization with Credibility Supervision}

\author{%
Liang Qian, Penggao Yan, Penghui Xu, and Li-Ta Hsu\thanks{Corresponding author: Li-Ta Hsu. Email: \href{mailto:lit.hsu@polyu.edu.hk}{lit.hsu@polyu.edu.hk}.}\\
\small Department of Aeronautical and Aviation Engineering\\
\small The Hong Kong Polytechnic University, Hong Kong, China
}

\date{}

\begin{document}

\maketitle

\begin{abstract}
Global navigation satellite system (GNSS) positioning is widely used for urban navigation, but the covariance reported by the GNSS solver is often unreliable in urban canyons. Existing differentiable factor graph optimization (DFGO) methods already learn measurement weighting through the solver, but they still use position-only objectives. As a result, the mean estimate may improve while the reported covariance remains too small, too large, or wrong in shape. In this work, we propose CredibleDFGO (CDFGO), a differentiable GNSS factor graph framework that makes covariance credibility an explicit training target. The Weighting Generation Network (WGN) predicts per-satellite reliability weights. The differentiable Gauss--Newton solver maps these weights to a position estimate and posterior covariance, and proper scoring rules supervise the East--North predictive distribution end-to-end. We study negative log-likelihood (NLL), Energy Score (ES), and their combination. Results on three UrbanNav test scenes show consistent gains in uncertainty credibility. Positioning accuracy also improves on the medium-urban and harsh-urban scenes, and the mean horizontal error and 95th-percentile error improve on the deep-urban scene. On the harsh-urban Mong Kok (MK) scene, CDFGO-Combined reduces the mean horizontal error from 13.77\,m to 11.68\,m, reduces NLL from 40.63 to 6.59, and reduces ES from 12.31 to 9.05. The case studies link the MK improvement to better axis-wise consistency, more credible local covariance ellipses, and satellite-level reweighting.
\end{abstract}


\section{Introduction}
\label{sec:intro}

GNSS positioning in urban canyons can return an accurate position while the reported covariance remains statistically inconsistent with the realized position-error distribution. This mismatch matters because the reported uncertainty is not merely a diagnostic output; it is used downstream. It affects how strongly fusion modules weight the GNSS position update and informs integrity-monitoring and alerting logic. A miscalibrated covariance can therefore bias both fusion weighting and integrity decisions. Li and colleagues formalized this consistency question as credibility \citep{li2012credibility}. Credibility assesses whether the reported uncertainty is statistically consistent with empirical position errors. Position accuracy alone does not tell us whether the reported uncertainty is credible. Yan and colleagues recently showed that two proper scoring rules, negative log-likelihood (NLL) and the energy score (ES), can detect different kinds of covariance miscalibration in state estimators under noise and model mismatch \citep{yan2025credibleuq}. In horizontal positioning, the relevant covariance failures are concrete: the reported ellipse may be too small, too large, or incorrectly oriented. A tight ellipse can still fail to cover the true East--North position. Two downstream settings show why this risk is not merely theoretical. Joerger and Pervan showed that sequential advanced receiver autonomous integrity monitoring (ARAIM) can exploit satellite motion over time only when the uncertainty model stays credible \citep{joerger2020araimmotion}. In driverless urban navigation, Nagai and colleagues found that fault-free integrity requires credible GNSS uncertainty for fusion with other onboard sensors \citep{nagai2024faultfreeintegrity}. Therefore, credibility cannot be inferred from the position error alone. It must be assessed from the reported covariance together with the realized errors. This requirement is not tied to one estimator family; it applies to classical weighted least squares, robust filters, and learned factor graphs.

Yet most work in urban GNSS targets position accuracy, not covariance credibility. Visibility and propagation-aware methods predict satellite visibility, distinguish LOS from NLOS signals, or use 3D-mapping-aided GNSS before the final position estimate is computed \citep{zheng2024visibility,wang2013shadowmatching,ng2021smartphone3dma}. Learning-based correction and weighting methods further improve estimator inputs or outputs by predicting NLOS-aware weights, smartphone GNSS corrections, adaptive stochastic models, pseudorange corrections, or position corrections from residual and line-of-sight features \citep{li2023nlosweighting,mohanty2023gcnncorrections,smolyakov2020multipathprediction,chen2023pseudorangecorrection,kanhere2022nncorrections}. Zhang and colleagues took a different direction by making measurement uncertainty itself the prediction target, using an LSTM to model its time-varying behavior \citep{zhang2021urbangnssuncertainty}. This, however, concerns measurement-level uncertainty rather than the Hessian-derived posterior covariance returned by the positioning solver. Robust-estimation studies have also shown that standard least squares becomes unreliable under outliers and heavy tails, and that urban M-estimator performance is sensitive to the assumed error model and estimator tuning \citep{medina2019robuststats,crespillo2020mestimators}. Huber tuning and multi-constellation fault isolation can suppress large errors \citep{li2026lah,yan2025multifaults}. Taken together, these methods mainly improve measurement selection, correction, weighting, measurement-level uncertainty modeling, or robust estimation. They do not treat the covariance returned by the solver as an explicit supervision target.

A similar gap persists within the factor-graph framework. Several methods address measurement faults and integrity monitoring, but they still do not directly supervise the solver covariance. Wen, Hsu, and colleagues brought factor-graph optimization into robust GNSS and RTK positioning \citep{wen2021icra}. They later applied graduated non-convexity to downweight pseudorange outliers \citep{wen2022gnc}, and a related thread from the same group fused GNSS and LiDAR with self-adaptive Gaussian mixtures \citep{wen2020gmmfusion}. Yan and colleagues learned subspace-based adaptive GMM error models for fault-aware urban positioning \citep{yan2025subspacegmm}. Knowles and Gao developed rapid fault detection and exclusion \citep{knowles2023edmfde}. For long-term integrity, Gallon and colleagues hardened orbit and clock error modeling \citep{gallon2022orbitclock}. Xia and colleagues came closest to the covariance credibility question: their integrity-constrained factor graph uses switch variables to reweight measurements and derives a protection level from the modified weighting matrix, though their method does not involve learning \citep{xia2024integrity}. Xia et al.'s work shows that factor graphs can incorporate covariance-related quantities in integrity analysis. For learned weighting, however, the solver covariance still needs an explicit training signal.

One might instead add generic uncertainty tools around the estimator, but this does not directly solve the covariance-calibration problem. Gal and Ghahramani's Monte Carlo dropout approximates Bayesian uncertainty by sampling subnetworks, whereas a factor-graph backend already exposes a Hessian-derived local covariance \citep{gal2016dropout}. For our problem, the issue is therefore not how to sample network uncertainty, but how to calibrate the covariance returned by the solver. Levi and colleagues showed how to calibrate regression uncertainty after training \citep{levi2022regressioncalibration}. Post-hoc calibration can adjust uncertainty after estimation, but it cannot alter the measurement weights that shape the forward solution. None of these approaches uses the realized position-error distribution to directly supervise the solver-returned covariance during end-to-end GNSS estimation.

These limitations motivate our route: supervising covariance within a differentiable factor graph. In this setting, the solver is unrolled so that gradients flow from the loss back through the optimization to the factor weights. Because the local posterior covariance is approximated by the inverse Hessian at the converged solution, a credibility loss can penalize a miscalibrated predictive distribution through the same backward pass.

The resulting framework builds on two lines of work: factor-graph smoothing for navigation and differentiable optimization for learning. Dellaert and Kaess formulated factor graphs and smoothing as a sparse probabilistic framework that incorporates measurements, priors, and constraints within one optimization problem \citep{dellaert2017factorgraphs}. The iSAM2 and Bayes tree formulations from Kaess and colleagues made incremental smoothing practical at navigation scale \citep{kaess2012isam2}, and Taylor and Gross recently framed factor-graph navigation as a generalization of Kalman filtering for asynchronous multi-sensor systems \citep{taylor2024factorgraphs}. Several groups have also integrated learning into GNSS estimators. Mohanty and Gao connected a graph neural network to a backpropagation Kalman filter so that the learned corrections directly affect the state estimate \citep{mohanty2024gnnkf}. Hu and colleagues coupled deep bias and weight prediction with RTKLIB-style positioning \citep{hu2025tdlgnss}. These works show that learned modules can be embedded in GNSS estimators, but they do not directly supervise the solver covariance. Yi and colleagues addressed differentiable factor graphs by unrolling the optimization steps and backpropagating through them \citep{yi2021dfgo}, establishing DFGO as a training paradigm. The idea of differentiating through a solved optimization problem is not limited to navigation. Amos and Kolter treated optimization as a neural-network layer in OptNet and derived exact gradients through solved programs \citep{amos2017optnet}. Xu and colleagues' DFGO-ICov brought this idea to smartphone GNSS. In DFGO-ICov, a network predicts adaptive measurement-factor covariance scales for pseudorange factors and is trained with a ground-truth position loss \citep{xu2023dfgo_icov}. Their follow-up AutoW replaced ground-truth labels with self-supervised priors from clustering and zero-velocity constraints, reporting gains over classical weighting schemes \citep{xu2024autow}. However, the training loss in these DFGO methods still acts on the position estimate, not on the Hessian-derived posterior covariance of the East--North solution. A position-only loss cannot distinguish a covariance that is sharp and statistically consistent from one that is sharp but misoriented, because both can yield the same position estimate.

Accordingly, CDFGO treats the solver output as a predictive distribution rather than as a point estimate alone. The training loss needs to evaluate both components of this distribution: the East--North mean and the posterior covariance. Proper scoring rules provide such a loss. In expectation over repeated samples, a strictly proper scoring rule attains its lowest value only when the predicted distribution matches the true conditional distribution. This means a covariance estimate that looks precise but is poorly calibrated is penalized, not rewarded. Axis-wise scores are useful diagnostics, but they do not fully capture East--North correlation \citep{gneiting2007,gneiting2008,scheuerer2015}. Yan and colleagues already used proper scoring rules as credibility metrics for state estimators \citep{yan2025credibleuq}. CDFGO uses these rules as training losses inside a differentiable solver. For the East--North Gaussian predictor used here, NLL penalizes a position estimate that is far from the true position and a covariance ellipse that is too small, too large, or incorrectly oriented. The energy score evaluates the predictive distribution through sample distances rather than through a closed-form Gaussian likelihood. In CDFGO, the predictive family remains the solver's Gaussian approximation, but ES provides a complementary multivariate training signal under heavy-tailed urban error conditions. For marginal axis-wise checks, a calibrated Gaussian predictor should place about 68.3\% and 99.7\% of realized errors within the reported $\pm1\sigma$ and $\pm3\sigma$ bounds, respectively.

We implement this approach as CredibleDFGO (CDFGO), an end-to-end framework for credibility supervision. A Weighting Generation Network (WGN) maps per-satellite features to reliability weights. A differentiable Gauss--Newton solver \citep{theseus2022} uses these values to set the per-factor information terms, and gradients flow through the solver back to the WGN parameters \citep{amos2017optnet}. We study three variants. CDFGO-NLL optimizes the Gaussian NLL of the East--North posterior, CDFGO-ES optimizes the energy score, and CDFGO-Combined uses both objectives. We evaluate CDFGO on three UrbanNav test scenes covering medium-urban, deep-urban, and harsh-urban conditions. Credibility-driven weighting may suppress contaminated satellites even when they improve nominal geometry. This can weaken the weighted geometry, but it can also produce a covariance that better reflects the realized positioning risk. We analyze this behavior as a geometry--credibility tradeoff.

The contributions of this paper are threefold:
\begin{itemize}
    \item We show that proper scoring rules, when used as end-to-end training losses for a differentiable GNSS factor graph, make the Hessian-derived posterior covariance an explicit supervised quantity rather than an unsupervised byproduct.
    \item We develop the CredibleDFGO framework and demonstrate on three UrbanNav scenes that credibility-driven training improves covariance credibility while preserving or improving overall horizontal accuracy and tail behavior.
    \item We identify and analyze a geometry--credibility tradeoff induced by credibility-driven weighting. A satellite-level analysis on the Mong Kok scene shows how this tradeoff reshapes the posterior covariance and helps explain the observed gains in both accuracy and credibility.
\end{itemize}

\section{Method}
\label{sec:method}

\subsection{Problem Setup and Notation}
\label{subsec:problem_setup}

We consider pseudorange-based GNSS positioning in urban environments. The corresponding state $\vect{x}_t$ is written as
\begin{equation}
\vect{x}_t =
\begin{bmatrix}
\vect{p}_t^\top & \vect{b}_t^\top
\end{bmatrix}^\top,
\qquad
\vect{b}_t =
\begin{bmatrix}
b_{t,\mathrm{G/J}} &
b_{t,\mathrm{E}} &
b_{t,\mathrm{R}} &
b_{t,\mathrm{C}}
\end{bmatrix}^{\top},
\end{equation}
where $\vect{p}_t = [E_t, N_t, U_t]^\top \in \mathbb{R}^3$ is the receiver position in the local East--North--Up frame. The vector $\vect{b}_t \in \mathbb{R}^4$ stores the constellation-specific clock biases in meters. Its first component is shared by GPS and QZSS. The remaining components correspond to Galileo, GLONASS, and BeiDou. Let $\mathcal{S}_t$ denote the set of visible satellites at epoch $t$. For each $s \in \mathcal{S}_t$, the observed pseudorange is modeled as
\begin{equation}
\rho_{t,s}^{\mathrm{obs}} = \rho_{t,s}^{\mathrm{pred}}(\vect{x}_t) + v_{t,s},
\label{eq:pseudorange_model_new}
\end{equation}
where $v_{t,s}$ is the measurement error. The pseudorange model follows the standard GNSS observation model \citep{kaplan2017gps}. The predicted pseudorange is
\begin{equation}
\rho_{t,s}^{\mathrm{pred}}(\vect{x}_t)
= \|\vect{p}_t - \vect{s}_{t,s}\|_2 + \vect{u}_{t,s}^{\top}\vect{b}_t + \Delta\rho_{t,s}^{\mathrm{corr}},
\label{eq:pseudorange_pred_new}
\end{equation}
where $\vect{s}_{t,s}$ is the satellite position, $\vect{u}_{t,s}\in\{0,1\}^{4}$ selects the clock-bias component for the constellation of satellite $s$, and $\Delta\rho_{t,s}^{\mathrm{corr}}$ collects the standard corrections, including satellite clock and atmospheric terms \citep{teunissen2017handbook}. Our aim is to estimate the position $\vect{\hat{p}}_t$ at each epoch through a learning-based approach, namely, CDFGO, as illustrated in Section~\ref{subsec:pipeline}.

\subsection{CDFGO Pipeline Overview}
\label{subsec:pipeline}

\begin{figure*}[!htb]
    \centering
    \includegraphicssafe[width=0.78\textwidth]{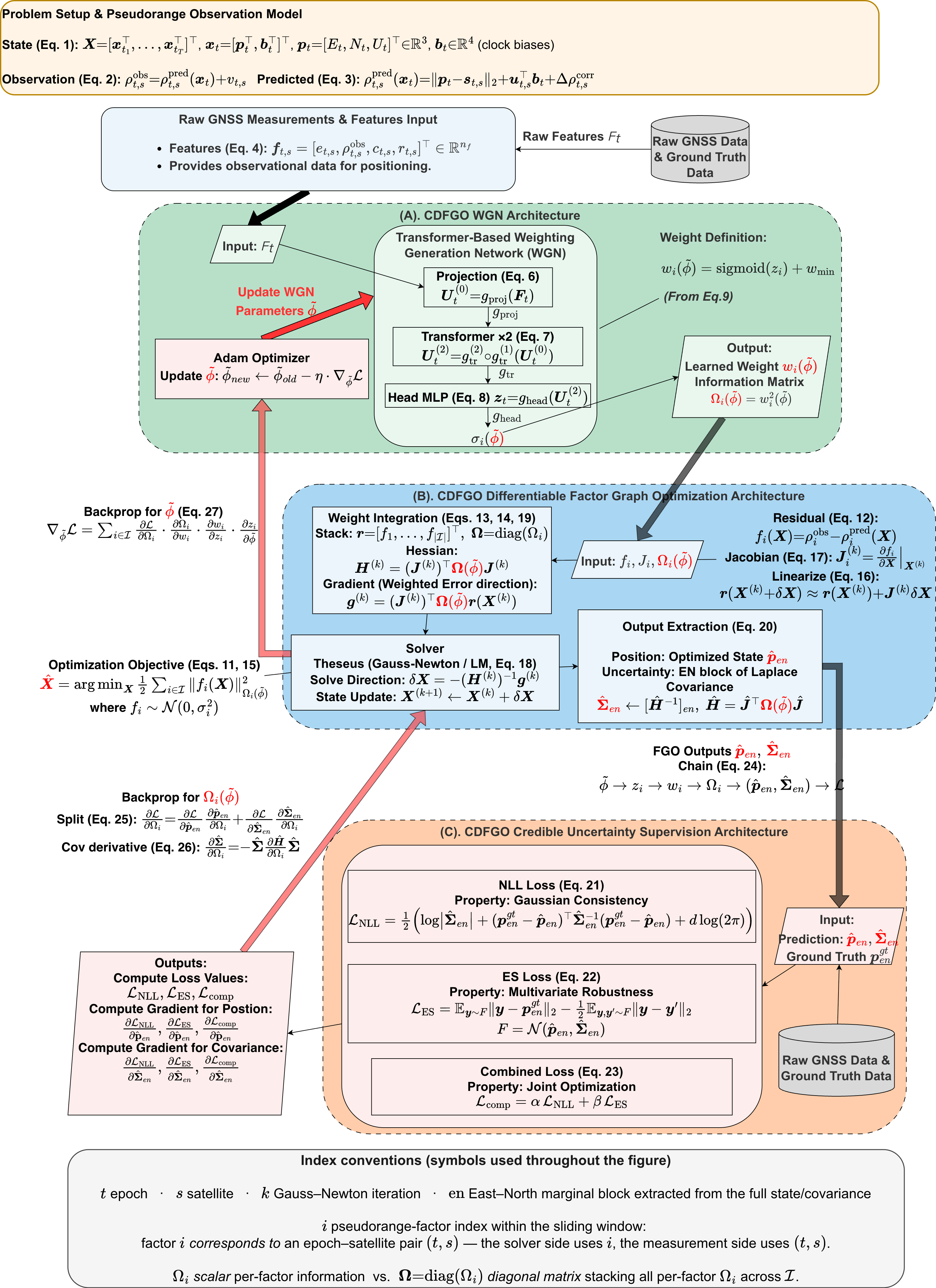}
    \caption{Overview of CDFGO. Panel (A) shows the WGN, which maps raw GNSS features to per-factor weights. Panel (B) shows the differentiable factor-graph solver, which uses these weights to produce state estimates and posterior covariance. Panel (C) shows the credibility supervision stage, where NLL/ES-based losses are back-propagated through the whole pipeline.}
    \label{fig:cdfgo_framework}
\end{figure*}

Figure~\ref{fig:cdfgo_framework} summarizes the CDFGO pipeline: the WGN, the differentiable Gauss--Newton solver, and credibility supervision.

During training, the WGN converts raw GNSS features into reliability weights. These weights define the pseudorange-factor information used by the differentiable Gauss--Newton solver, which returns the state estimate and posterior covariance. Credibility losses are then computed from the East--North position and covariance returned by the solver, and the loss gradients are back-propagated through the solver to the WGN parameters $\tilde{\phi}$.

During inference, only the trained WGN and the solver are used. The runtime solver therefore stays the same as in a standard FGO backend; only the factor weights are learned from data.

\FloatBarrier

\subsection{Weighting Generation Network}
\label{subsec:wgn_new}

For each visible satellite $s \in \mathcal{S}_t$, the feature vector $\vect{f}_{t,s}$ contains four inputs: elevation angle $e_{t,s}$, observed pseudorange $\rho_{t,s}^{\mathrm{obs}}$, carrier-to-noise ratio $c_{t,s}$, and pseudorange residual $r_{t,s}$. Here $r_{t,s}$ is the residual of a preliminary pseudorange-only weighted least-squares estimate, with weights from the GoGPS scheme \citep{realini2013gogps}. This auxiliary solution provides a lightweight fit-based feature for each satellite before the main factor-graph solver runs. Stacking the satellite features gives the input matrix $\mat{F}_t$.
\begin{equation}
\vect{f}_{t,s} =
\begin{bmatrix}
e_{t,s} &
\rho_{t,s}^{\mathrm{obs}} &
c_{t,s} &
r_{t,s}
\end{bmatrix}^{\top}
\in \mathbb{R}^{n_f},
\end{equation}
\begin{equation}
\mat{F}_t =
\begin{bmatrix}
\vect{f}_{t,s_1}^{\top} \\
\vect{f}_{t,s_2}^{\top} \\
\vdots \\
\vect{f}_{t,s_{|\mathcal{S}_t|}}^{\top}
\end{bmatrix}
\in \mathbb{R}^{|\mathcal{S}_t| \times n_f}.
\end{equation}

As shown in Fig.~\ref{fig:cdfgo_framework}(A), a projection MLP first maps $\mat{F}_t$ to latent vectors $\mat{U}_t^{(0)}$. Two transformer encoder layers \citep{vaswani2017} then refine these vectors to $\mat{U}_t^{(1)}$ and $\mat{U}_t^{(2)}$, allowing each satellite representation to incorporate the other visible satellites at the same epoch. A final MLP head outputs the per-factor scores $\vect{z}_t$ from $\mat{U}_t^{(2)}$:
\begin{equation}
\mat{U}_t^{(0)} = g_{\mathrm{proj}}(\mat{F}_t),
\end{equation}
\begin{equation}
\mat{U}_t^{(1)} = g_{\mathrm{tr}}^{(1)}(\mat{U}_t^{(0)}), \qquad
\mat{U}_t^{(2)} = g_{\mathrm{tr}}^{(2)}(\mat{U}_t^{(1)}),
\end{equation}
\begin{equation}
\vect{z}_t = \mathrm{WGN}(\mat{F}_t;\tilde{\phi}) = g_{\mathrm{head}}(\mat{U}_t^{(2)}),
\end{equation}
where $\tilde{\phi}$ denotes the WGN parameters.

For factor $i$, the scalar output $z_i$ is mapped to a positive reliability weight $w_i(\tilde{\phi})$ through a shifted sigmoid that ensures $w_i(\tilde{\phi}) \ge w_{\min}$. 
\begin{equation}
w_i(\tilde{\phi}) = \mathrm{sigmoid}(z_i) + w_{\min},
\label{eq:weight_def_new}
\end{equation}
The corresponding standard deviation $\sigma_i$ and scalar information term $\Omega_i$ used by the factor graph solver are then
\begin{equation}
\sigma_i(\tilde{\phi}) = \frac{1}{w_i(\tilde{\phi})}, \qquad
\Omega_i(\tilde{\phi}) = \frac{1}{\sigma_i^2(\tilde{\phi})} = w_i^2(\tilde{\phi}).
\label{eq:sigma_omega_new}
\end{equation}
These $\Omega_i$ values are the only learned quantities passed from the WGN to the solver.

\subsection{Differentiable FGO and Posterior Covariance}
\label{subsec:dfgo_new}

The solver stage, shown in Fig.~\ref{fig:cdfgo_framework}(B), follows the standard maximum a posteriori (MAP) factor-graph formulation \citep{dellaert2017factorgraphs,kschischang2001factor}. Let $\vect{X}$ denote the stacked state vector over a short sliding window of consecutive epochs, and let $\mathcal{I}$ denote the corresponding set of pseudorange factors in that local batch; the specific window length is given in Section~\ref{sec:experiments_new}. Each epoch contributes its own state and pseudorange factors, and the sliding window is used only to form a local batch; no additional inter-epoch factor is introduced. Given the factor weights $\Omega_i$ from the WGN, the state $\hat{\vect{X}}$ is obtained by solving
\begin{equation}
\hat{\vect{X}} =
\argmin_{\vect{X}}
\frac{1}{2}
\sum_{i\in\mathcal{I}}
\left\| f_i(\vect{X}) \right\|_{\Omega_i(\tilde{\phi})}^{2},
\label{eq:map_new}
\end{equation}
where $f_i(\vect{X})$ is the residual of factor $i$ and $\|r\|_{\Omega}^{2}=r^{\top}\Omega r$. For scalar pseudorange factors, the information term $\Omega_i(\tilde{\phi})$ is also scalar and follows Eq.~\eqref{eq:sigma_omega_new}. The residual of factor $i$ is
\begin{equation}
f_i(\vect{X}) = \rho_i^{\mathrm{obs}} - \rho_i^{\mathrm{pred}}(\vect{X}),
\label{eq:residual_new}
\end{equation}
where $\rho_i^{\mathrm{obs}}$ is the observed pseudorange and $\rho_i^{\mathrm{pred}}(\vect{X})$ is the predicted pseudorange. The predicted term consists of the geometric range, the relevant constellation clock term, and the corrections defined in Eq.~\eqref{eq:pseudorange_pred_new}. Stacking all factor residuals gives the residual vector $\vect{r}(\vect{X})$, and stacking all scalar information terms gives the diagonal information matrix $\mat{\Omega}$:
\begin{equation}
\vect{r}(\vect{X}) =
\begin{bmatrix}
f_1(\vect{X}) &
f_2(\vect{X}) &
\cdots &
f_{|\mathcal{I}|}(\vect{X})
\end{bmatrix}^{\top},
\end{equation}
\begin{equation}
\mat{\Omega} =
\mathrm{diag}\!\left(\Omega_1,\Omega_2,\ldots,\Omega_{|\mathcal{I}|}\right).
\end{equation}
Then Eq.~\eqref{eq:map_new} can be written in matrix form as
\begin{equation}
\hat{\vect{X}} =
\argmin_{\vect{X}}
\frac{1}{2}
\vect{r}(\vect{X})^{\top}
\mat{\Omega}
\vect{r}(\vect{X}).
\end{equation}

This nonlinear least-squares problem is solved by iterative linearization. At iteration $k$, around the current estimate $\vect{X}^{(k)}$, the residual vector is approximated by its first-order expansion. The Jacobian matrix $\mat{J}^{(k)}$ collects the derivatives of all residuals with respect to the state, and its $i$-th row $\vect{J}_i^{(k)}$ corresponds to factor $i$:
\begin{equation}
\vect{r}(\vect{X}^{(k)}+\delta\vect{X})
\approx
\vect{r}(\vect{X}^{(k)}) + \mat{J}^{(k)} \delta\vect{X},
\end{equation}
\begin{equation}
\mat{J}^{(k)} =
\frac{\partial \vect{r}}{\partial \vect{X}}
\bigg|_{\vect{X}^{(k)}},
\qquad
\vect{J}_i^{(k)} =
\frac{\partial f_i}{\partial \vect{X}}
\bigg|_{\vect{X}^{(k)}}.
\end{equation}
Since $f_i(\vect{X})=\rho_i^{\mathrm{obs}}-\rho_i^{\mathrm{pred}}(\vect{X})$, the Jacobian is obtained by differentiating the predicted pseudorange model with respect to the state variables. Substituting the linearized residual into the objective gives a local quadratic approximation. This leads to the standard Gauss--Newton normal equation \citep{dellaert2017factorgraphs,kaess2012isam2}
\begin{equation}
\mat{H}^{(k)} \delta\vect{X} = -\vect{g}^{(k)},
\end{equation}
with the gradient term $\vect{g}^{(k)}$ and the Hessian approximation $\mat{H}^{(k)}$ given by
\begin{equation}
\vect{g}^{(k)} =
(\mat{J}^{(k)})^{\top}\mat{\Omega}\,\vect{r}(\vect{X}^{(k)}),
\qquad
\mat{H}^{(k)} =
(\mat{J}^{(k)})^{\top}\mat{\Omega}\,\mat{J}^{(k)}.
\label{eq:hessian_new}
\end{equation}
The iterative solver uses a differentiable Gauss--Newton scheme implemented in the Theseus differentiable optimization library \citep{yi2021dfgo,theseus2022}.

After convergence, the solver returns the state estimate $\hat{\vect{X}}$ and the posterior covariance \citep{dellaert2017factorgraphs}, as indicated in Fig.~\ref{fig:cdfgo_framework}(B). The posterior covariance $\hat{\mat{\Sigma}}$ is approximated by the inverse of the Hessian approximation $\hat{\mat{H}}$ evaluated at the converged state $\hat{\vect{X}}$,
\begin{equation}
\hat{\mat{\Sigma}} \approx \hat{\mat{H}}^{-1},
\label{eq:laplace_new}
\end{equation}
which is the standard covariance approximation around the converged solution in nonlinear least-squares factor-graph estimation \citep{dellaert2017factorgraphs}. From the block of $\hat{\vect{X}}$, the East--North position estimate $\hat{\vect{p}}_{en}$ is extracted. From $\hat{\mat{\Sigma}}$, the East--North marginal covariance $\hat{\mat{\Sigma}}_{en}$ is extracted. This horizontal block $\hat{\mat{\Sigma}}_{en}$ is used because the credibility objectives are defined on horizontal positioning.

\subsection{Credibility Objectives}
\label{subsec:loss_new}

Three credibility objectives are used to train the WGN: the negative log-likelihood (NLL), the Energy Score (ES), and their combination. Each evaluates the EN predictive distribution defined by the solver's position estimate $\hat{\vect{p}}_{en}$ and covariance $\hat{\mat{\Sigma}}_{en}$ against the ground-truth EN position $\vect{p}^{gt}_{en}$.

The Laplace approximation in Eq.~\eqref{eq:laplace_new} implies a Gaussian predictive distribution $\mathcal{N}(\hat{\vect{p}}_{en},\hat{\mat{\Sigma}}_{en})$ for the EN position \citep{dellaert2017factorgraphs}. The first objective is the negative log-likelihood (NLL) of this Gaussian at the ground-truth position, a standard strictly proper scoring rule for Gaussian forecasts \citep{gneiting2007}:
\begin{equation}
\mathcal{L}_{\mathrm{NLL}}=
\frac{1}{2}\left[
\log|\hat{\mat{\Sigma}}_{en}|+
\left(\vect{p}^{gt}_{en}-\hat{\vect{p}}_{en}\right)^\top\hat{\mat{\Sigma}}_{en}^{-1}\left(\vect{p}^{gt}_{en}-\hat{\vect{p}}_{en}\right)
+d\log(2\pi)
\right],
\label{eq:nll_new}
\end{equation}
where $d=2$ is the dimension of the EN prediction. The quadratic term $(\vect{p}^{gt}_{en}-\hat{\vect{p}}_{en})^\top \hat{\mat{\Sigma}}_{en}^{-1} (\vect{p}^{gt}_{en}-\hat{\vect{p}}_{en})$ measures the position error relative to the predicted covariance, while the log-determinant term $\log|\hat{\mat{\Sigma}}_{en}|$ penalizes unnecessarily large covariance. NLL therefore encourages both an accurate mean estimate and a covariance that matches the realized error magnitude.

The second objective is the multivariate Energy Score (ES), defined in Eq.~\eqref{eq:es_new}:
\begin{equation}
\mathcal{L}_{\mathrm{ES}}
= \mathbb{E}_{\vect{y}\sim F}\|\vect{y}-\vect{p}^{gt}_{en}\|_2
-\frac{1}{2}\mathbb{E}_{\vect{y},\vect{y}'\sim F}\|\vect{y}-\vect{y}'\|_2,
\label{eq:es_new}
\end{equation}
where $F=\mathcal{N}(\hat{\vect{p}}_{en},\hat{\mat{\Sigma}}_{en})$ is the same predictive distribution as in NLL. The first term $\mathbb{E}_{\vect{y}\sim F}\|\vect{y}-\vect{p}^{gt}_{en}\|_2$ measures how far predictive samples are from the ground truth, and the second term $\frac{1}{2}\mathbb{E}_{\vect{y},\vect{y}'\sim F}\|\vect{y}-\vect{y}'\|_2$ measures the dispersion of the predictive distribution. ES is used here because it evaluates the predictive distribution through distances to the ground truth and between predictive samples \citep{gneiting2008}. During training, $\mathcal{L}_{\mathrm{ES}}$ is estimated with Monte Carlo samples from $F$ using the reparameterization trick \citep{kingma2014vae,rezende2014stochastic}.

The combined objective is
\begin{equation}
\mathcal{L}_{\mathrm{comp}}=\alpha\mathcal{L}_{\mathrm{NLL}}+\beta\mathcal{L}_{\mathrm{ES}},
\label{eq:comp_new}
\end{equation}
where $\alpha\ge 0$ and $\beta\ge 0$ balance the two terms. The combined objective applies both the likelihood-based penalty from NLL and the sample-based distance score from ES.

\subsection{Closed-Loop Differentiable Training Process}
\label{subsec:training_process_new}

Let $\mathcal{L}$ denote any one of $\mathcal{L}_{\mathrm{NLL}}$, $\mathcal{L}_{\mathrm{ES}}$, or $\mathcal{L}_{\mathrm{comp}}$. Training the WGN requires gradients of $\mathcal{L}$ with respect to the WGN parameters $\tilde{\phi}$. Because $\mathcal{L}$ depends on the solver outputs $\hat{\vect{p}}_{en}$ and $\hat{\mat{\Sigma}}_{en}$, the gradient must pass through the differentiable solver. The forward dependency chain from the parameters to the loss is
\begin{equation*}
\tilde{\phi}\;\longrightarrow\;z_i\;\longrightarrow\;w_i\;\longrightarrow\;\Omega_i\;\longrightarrow\;(\hat{\vect{p}}_{en},\;\hat{\mat{\Sigma}}_{en})\;\longrightarrow\;\mathcal{L},
\end{equation*}
and the gradient flows in the opposite direction during back-propagation.

Since both $\hat{\vect{p}}_{en}$ and $\hat{\mat{\Sigma}}_{en}$ depend on the information value $\Omega_i$ assigned to factor $i$, the gradient splits into two additive paths, as shown in Eq.~\eqref{eq:grad_omega_new}:
\begin{equation}
\frac{\partial\mathcal{L}}{\partial\Omega_i}
=
\underbrace{
\frac{\partial\mathcal{L}}{\partial\hat{\vect{p}}_{en}}
\frac{\partial\hat{\vect{p}}_{en}}{\partial\Omega_i}
}_{\text{position path}}
+
\underbrace{
\frac{\partial\mathcal{L}}{\partial\hat{\mat{\Sigma}}_{en}}
\frac{\partial\hat{\mat{\Sigma}}_{en}}{\partial\Omega_i}
}_{\text{covariance path}}.
\label{eq:grad_omega_new}
\end{equation}

In the position path, $\Omega_i$ controls how strongly each measurement contributes to the solution. It enters the Hessian $\mat{H}^{(k)}$ and the gradient vector $\vect{g}^{(k)}$ in Eq.~\eqref{eq:hessian_new}. These two quantities determine the Gauss--Newton update $\delta\vect{X}=-(\mat{H}^{(k)})^{-1}\vect{g}^{(k)}$. Repeating this update produces the state estimate $\hat{\vect{X}}$, from which $\hat{\vect{p}}_{en}$ is extracted. Changing $\Omega_i$ therefore changes the position estimate and the loss.

In the covariance path, the derivation first differentiates the full posterior covariance $\hat{\mat{\Sigma}}\approx\hat{\mat{H}}^{-1}$ from Eq.~\eqref{eq:laplace_new} and then extracts the East--North block $\hat{\mat{\Sigma}}_{en}$. Changing $\Omega_i$ changes the final Hessian and therefore changes both the full posterior covariance and the derived EN block. Applying the inverse-matrix derivative identity gives
\begin{equation}
\frac{\partial\hat{\mat{\Sigma}}}{\partial\Omega_i}
=
-\hat{\mat{\Sigma}}\,\frac{\partial\hat{\mat{H}}}{\partial\Omega_i}\,\hat{\mat{\Sigma}},
\label{eq:cov_grad_new}
\end{equation}
where $\partial\hat{\mat{H}}/\partial\Omega_i$ denotes the total sensitivity of the final Hessian to $\Omega_i$, including the explicit weight term in $\mat{J}^{\top}\mat{\Omega}\mat{J}$ and the indirect dependence on the converged state. This indirect dependence remains inside the covariance-path term $\partial\hat{\mat{\Sigma}}_{en}/\partial\Omega_i$ in Eq.~\eqref{eq:grad_omega_new}; it does not introduce a third top-level loss path. In practice, these terms are obtained by back-propagating through the unrolled differentiable solver \citep{yi2021dfgo,theseus2022}.

After computing $\partial\mathcal{L}/\partial\Omega_i$, Eqs.~\eqref{eq:sigma_omega_new} and \eqref{eq:weight_def_new} provide $\partial\Omega_i/\partial w_i$ and $\partial w_i/\partial z_i$. The final factor $\partial z_i/\partial\tilde{\phi}$ is the WGN Jacobian. Summing over all factors gives the parameter gradient in Eq.~\eqref{eq:grad_phi_new}
\begin{equation}
\nabla_{\tilde{\phi}}\mathcal{L}
=
\sum_{i\in\mathcal{I}}
\frac{\partial\mathcal{L}}{\partial\Omega_i}
\cdot\frac{\partial\Omega_i}{\partial w_i}
\cdot\frac{\partial w_i}{\partial z_i}
\cdot\frac{\partial z_i}{\partial\tilde{\phi}}.
\label{eq:grad_phi_new}
\end{equation}
The credibility loss therefore supervises the learned weights through both the position estimate and the covariance estimate.

\section{Experiments and Mechanism Analysis}
\label{sec:experiments_new}
\subsection{Dataset and Experiment Setup}
\label{subsec:dataset}

\begin{figure*}[!htb]
    \centering
    \begin{subfigure}[t]{0.32\textwidth}
        \centering
        \includegraphicssafe[width=\linewidth]{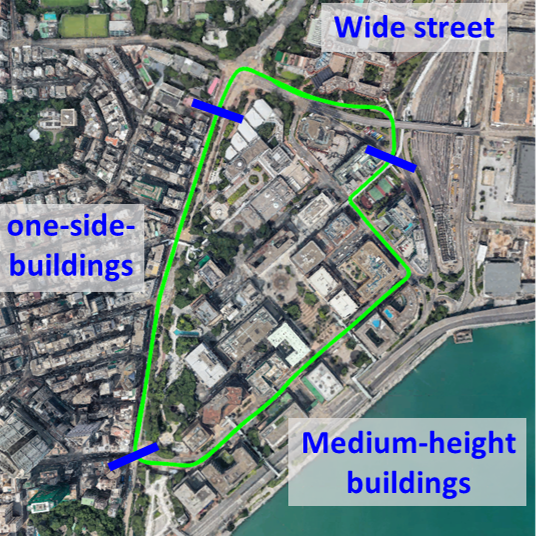}
        \caption{TST (Medium Urban).}
    \end{subfigure}
    \hfill
    \begin{subfigure}[t]{0.32\textwidth}
        \centering
        \includegraphicssafe[width=\linewidth]{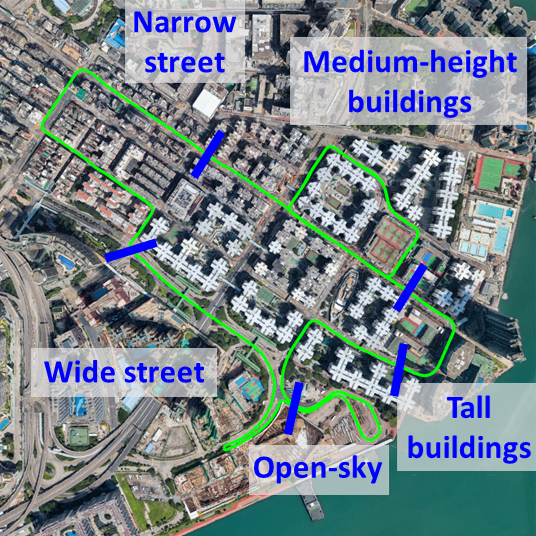}
        \caption{WP (Deep Urban).}
    \end{subfigure}
    \hfill
    \begin{subfigure}[t]{0.32\textwidth}
        \centering
        \includegraphicssafe[width=\linewidth]{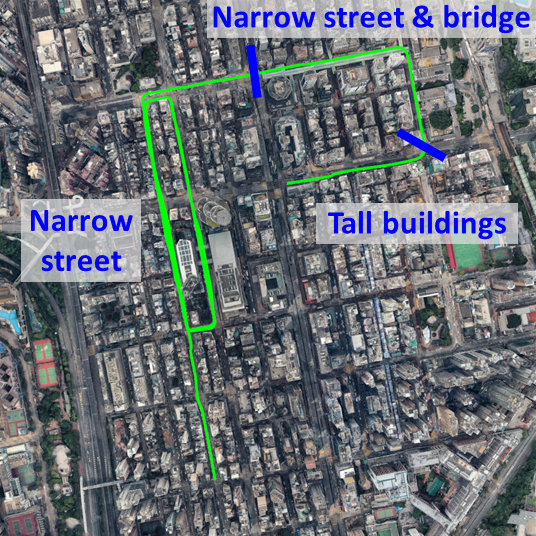}
        \caption{MK (Harsh Urban).}
    \end{subfigure}
    \caption{Scene overview of the three test scenes. The figures are reproduced from the UrbanNav dataset \citep{hsu2021urbannav,hsu2023urbannav}.}
    \label{fig:urbannav_test_scenes}
\end{figure*}

In this work, we use the UrbanNav dataset \citep{hsu2023urbannav}, which is a public multi-sensor urban-navigation dataset in Hong Kong covering medium- to harsh-urban scenes with RTK/INS reference trajectories, to evaluate the performance of the proposed method. Specifically, three test scenes are used: Tsim Sha Tsui (TST), Whampoa (WP), and Mong Kok (MK), representing Medium Urban, Deep Urban, and Harsh Urban conditions, respectively. Figure~\ref{fig:urbannav_test_scenes} shows aerial overviews of the three test scenes, illustrating how building density and street canyon depth increase from TST (medium urban) to WP (deep urban) to MK (harsh urban). All recordings use the same hardware setup, with a u-blox ZED-F9P receiver for raw GNSS measurements and a NovAtel SPAN-CPT RTK/INS system for the reference trajectory.
\FloatBarrier

The data are split by date and route subset to avoid temporal leakage between training and testing. Training uses Whampoa (WP), Integrated, and Kowloon Bay (KLB). Testing uses Tsim Sha Tsui (TST), Whampoa (WP), and Mong Kok (MK). Because the WP training and testing data were collected on different dates, WP serves as a temporal-transfer case. TST and MK are unseen routes and therefore serve as spatial-transfer test cases. Table~\ref{tab:dataset_summary} summarizes the scene type and epoch count of each split, with 15,507 training epochs and 4,373 testing epochs in total. No testing trajectory is used for feature normalization, parameter fitting, or model selection.

\begin{table*}[!htb]
    \centering
    \caption{Dataset used in this work.}
    \label{tab:dataset_summary}
    \footnotesize
    \begin{tabular}{@{}lllllr@{}}
        \toprule
        Set & ID & Date & Location & Scene & Epochs \\
        \midrule
        Train & 1 & 2021-07-14 & Whampoa (WP) & Deep Urban & 1,209 \\
        Train & 2 & 2021-08-17 & Integrated & Mixed Urban & 12,050 \\
        Train & 3 & 2021-12-03 & Kowloon Bay (KLB) & Deep Urban & 2,248 \\
        \multicolumn{5}{r}{Total training epochs} & 15,507 \\
        \midrule
        Test & 4 & 2021-05-17 & Tsim Sha Tsui (TST) & Medium Urban & 558 \\
        Test & 5 & 2021-05-21 & Whampoa (WP) & Deep Urban & 1,532 \\
        Test & 6 & 2021-05-18 & Mong Kok (MK) & Harsh Urban & 2,283 \\
        \multicolumn{5}{r}{Total testing epochs} & 4,373 \\
        \bottomrule
    \end{tabular}
\end{table*}

The WGN uses the four per-satellite features defined in Section~\ref{subsec:wgn_new}: elevation angle, observed pseudorange, carrier-to-noise ratio, and pseudorange residual. The residual feature is computed during data preparation from that preliminary WLS. All continuous inputs are normalized with training-set statistics only, and variable satellite counts are handled by padding and masking. The WGN is trained end-to-end with Adam \citep{kingma2014adam}, and its MLP layers use LeakyReLU activations. In the experiment, the weight lower bound is set to $w_{\min}=0$. The ES objective is estimated with Monte Carlo reparameterization using $K=2048$ samples per batch. For the combined objective, $\alpha=\beta=0.5$. The differentiable Gauss--Newton solver is implemented with Theseus \citep{theseus2022}, and all reported results use the same pseudorange-only solver setting with a fixed five-epoch sliding time window.

\subsection{Metrics, Baselines, and Variants}
\label{subsec:metrics_baselines}

In this experiment, point accuracy is evaluated with three horizontal-error metrics in meters: the mean horizontal error, the median (50th percentile), and the 95th percentile. The mean reflects the overall operating level, the median the typical-case performance, and the 95th percentile the tail behavior.

Uncertainty credibility is reported by two proper scoring rules, namely NLL and ES, both computed from the predicted EN Gaussian distribution $\mathcal{N}(\hat{\vect{p}}_{en},\hat{\mat{\Sigma}}_{en})$. NLL evaluates whether the reported covariance is statistically consistent with the realized error under the Gaussian predictive model \citep{gneiting2007}. ES evaluates distribution-level mismatch through sample distances \citep{gneiting2008,scheuerer2015}. Table~\ref{tab:metrics_summary} summarizes these five metrics and how each is interpreted in the present study.

\begin{table}[!htb]
    \centering
    \caption{Summary of evaluation metrics and their interpretation.}
    \label{tab:metrics_summary}
    \footnotesize
    \renewcommand{\arraystretch}{1.12}
    \setlength{\tabcolsep}{4pt}
    \begin{tabularx}{\linewidth}{@{}ll>{\raggedright\arraybackslash}X>{\raggedright\arraybackslash}X@{}}
        \toprule
        Category & Metric & Definition / Scope & Interpretation in This Study \\
        \midrule
        \multirow{3}{*}{Point Accuracy}
        & Mean & Mean 2D horizontal error over the full data. & Overall operating-level accuracy. \\
        & 50\% & Median 2D horizontal error. & Typical-case accuracy summarized by the median. \\
        & 95\% & 95th-percentile 2D horizontal error. & Tail behavior under difficult epochs. \\
        \midrule
        \multirow{2}{*}{Uncertainty Credibility}
        & NLL & Negative log-likelihood of the predicted EN Gaussian posterior evaluated at ground truth. & Penalizes mismatch between reported covariance and realized error under the Gaussian model. \\
        & ES & Multivariate energy score of the predictive distribution versus ground truth. & Measures distribution-level mismatch between prediction and outcome. \\
        \bottomrule
    \end{tabularx}
\end{table}

We compared the proposed method with both classical baselines and learning-based baselines. The classical baselines are the Elevation Model (elevation-angle-based weighting) \citep{teunissen2017handbook}, Sigma-$\epsilon$ \citep{hartinger1999sigma}, and the GoGPS weighting scheme \citep{realini2013gogps}. The learning baseline is DFGO (MAE), which trains the same differentiable solver with an MAE objective \citep{xu2023dfgo_icov}. The proposed variants are CDFGO-NLL, CDFGO-ES, and CDFGO-Combined. These variants differ only in the training objective and therefore isolate the effect of credibility supervision.

All methods use the same preprocessing pipeline, the same differentiable Gauss--Newton solver backend, and the same solver convergence criteria. The learning-based methods additionally share the same mini-batch schedule. Under this controlled setting, the comparison focuses on how changing the training objective affects both point accuracy and covariance credibility.

\subsection{Main Quantitative Results}
\label{subsec:main_results}

\begin{table*}[!htb]
    \centering
    \caption{Quantitative comparison of positioning accuracy (Mean, 50\%, and 95\% 2D error in meters) and uncertainty credibility (NLL and ES) across three urban scenes. Best performance in each column is highlighted in bold.}
    \label{tab:main_results}
    \footnotesize
    \setlength{\tabcolsep}{3.5pt}
    \renewcommand{\arraystretch}{1.08}
    \resizebox{\textwidth}{!}{%
    \begin{tabular}{@{}lccccc@{\hspace{8pt}}ccccc@{\hspace{8pt}}ccccc@{}}
        \toprule
        \multirow{2}{*}{Method}
        & \multicolumn{5}{c}{Medium Urban (TST)}
        & \multicolumn{5}{c}{Deep Urban (WP)}
        & \multicolumn{5}{c}{Harsh Urban (MK)} \\
        \cmidrule(lr){2-6} \cmidrule(lr){7-11} \cmidrule(l){12-16}
        & Mean & 50\% & 95\% & NLL & ES
        & Mean & 50\% & 95\% & NLL & ES
        & Mean & 50\% & 95\% & NLL & ES \\
        \midrule
        \multicolumn{16}{@{}l}{\textit{Baseline approaches}} \\
        Elevation Model
        & 10.82 & 7.58 & 36.76 & $4.5\times10^{3}$ & 10.69
        & 10.07 & 6.73 & 28.62 & $4.1\times10^{3}$ & 9.95
        & 20.88 & 15.55 & 53.57 & $1.6\times10^{4}$ & 20.72 \\
        Sigma-$\epsilon$
        & 8.06 & 5.19 & 26.96 & $8.4\times10^{5}$ & 8.05
        & 7.80 & 5.13 & 22.34 & $4.4\times10^{5}$ & 7.79
        & 13.95 & 10.39 & 36.47 & $2.2\times10^{6}$ & 13.94 \\
        GoGPS
        & 7.73 & 5.40 & 20.02 & 9.90 & 6.17
        & 7.81 & 5.41 & 21.13 & 8.76 & 6.39
        & 15.22 & 12.74 & 35.50 & 34.18 & 13.35 \\
        DFGO (MAE)
        & 6.27 & 4.27 & 19.31 & 14.06 & 5.23
        & 6.99 & \textbf{4.65} & 22.86 & 13.08 & 5.92
        & 13.77 & 10.04 & 36.55 & 40.63 & 12.31 \\
        \midrule
        \multicolumn{16}{@{}l}{\textit{Proposed CredibleDFGO variants}} \\
        CDFGO-NLL
        & 4.93 & 3.88 & 12.16 & 3.69 & 3.52
        & 6.78 & 4.98 & \textbf{17.29} & 4.11 & 4.92
        & 12.06 & \textbf{7.60} & 33.88 & 7.03 & 9.39 \\
        CDFGO-ES
        & \textbf{4.71} & \textbf{3.70} & \textbf{11.86} & 4.40 & \textbf{3.44}
        & 6.63 & 4.92 & 19.44 & 5.44 & 5.13
        & 12.01 & 7.70 & 34.71 & 10.75 & 9.70 \\
        CDFGO-Combined
        & 5.08 & 3.85 & 12.00 & \textbf{3.67} & 3.58
        & \textbf{6.44} & 4.84 & 17.39 & \textbf{3.76} & \textbf{4.61}
        & \textbf{11.68} & 7.89 & \textbf{33.26} & \textbf{6.59} & \textbf{9.05} \\
        \bottomrule
    \end{tabular}%
    }
\end{table*}

Table~\ref{tab:main_results} compares point accuracy and covariance credibility across the four baselines and three CredibleDFGO variants on the TST, WP, and MK scenes. On the medium-urban TST scene, all three CredibleDFGO variants outperform DFGO (MAE) in both point accuracy and credibility. CDFGO-ES gives the lowest mean horizontal error at 4.71\,m, compared with 6.27\,m for DFGO (MAE), and also gives the lowest 95th-percentile error at 11.86\,m. CDFGO-Combined gives the lowest NLL, reducing it from 14.06 to 3.67, a 73.9\% reduction. The three credibility-driven variants also reduce ES by similar margins, which shows that credibility supervision improves distribution-level performance on this medium-urban scene.

On the deep-urban WP scene, the pattern is different. DFGO (MAE) gives the lowest median error, so credibility-driven training does not improve the median. The main accuracy gain appears in the tail: CDFGO-Combined reduces the 95th-percentile error from 22.86\,m to 17.39\,m, a 23.9\% reduction, and lowers the mean horizontal error from 6.99\,m to 6.44\,m. The credibility metrics also improve, with NLL reduced from 13.08 to 3.76 and ES reduced from 5.92 to 4.61. CDFGO-NLL gives a slightly lower 95th-percentile error at 17.29\,m. On WP, credibility-driven training mainly improves tail behavior and covariance credibility rather than the median error.

On the harsh-urban MK scene, the separation between DFGO (MAE) and the credibility-driven objectives is largest. CDFGO-Combined reduces the mean horizontal error from 13.77\,m to 11.68\,m, a 15.2\% reduction, and reduces NLL from 40.63 to 6.59. ES also drops from 12.31 to 9.05, and the 95th-percentile error drops from 36.55\,m to 33.26\,m. The median error improves from 10.04\,m to 7.89\,m, while CDFGO-NLL gives the lowest median at 7.60\,m. Because MK shows the largest NLL reduction and joint gains in accuracy and credibility in Table~\ref{tab:main_results}, the mechanism analysis below focuses on DFGO (MAE) and CDFGO-Combined on this scene.

\subsection{Axis-wise Credibility Diagnostics on MK}
\label{subsec:credibility_diagnostics}

We compare the reported East--North uncertainty with the realized East--North errors on MK using the two diagnostics introduced by \citet{li2012credibility}. The first is the exceedance probability in Eq.~\eqref{eq:exceedance}, which measures the fraction of epochs whose absolute error exceeds a reported $k\hat{\sigma}$ bound. The second is in-envelope coverage in Eq.~\eqref{eq:coverage}, which measures the fraction of epochs whose absolute error remains within the same bound. For axis $d \in \{E,N\}$, these diagnostics are defined as
\begin{equation}
\mathbb{P}\!\left(|e_d| > k\hat{\sigma}_d\right)
=
\frac{1}{T}\sum_{t=1}^{T}
\mathbf{1}\!\left(|e_{d,t}| > k\hat{\sigma}_{d,t}\right),
\label{eq:exceedance}
\end{equation}
\begin{equation}
\mathbb{P}\!\left(|e_d| \le k\hat{\sigma}_d\right)
=
\frac{1}{T}\sum_{t=1}^{T}
\mathbf{1}\!\left(|e_{d,t}| \le k\hat{\sigma}_{d,t}\right),
\label{eq:coverage}
\end{equation}
where $e_{d,t}$ is the realized error on axis $d$ at epoch $t$, $\hat{\sigma}_{d,t}$ is the reported standard deviation on that axis, and $T$ is the number of epochs. For a Gaussian predictor, the nominal 3$\sigma$ exceedance is 0.27\%, and the nominal 1$\sigma$ in-envelope coverage is 68.27\%. These values are used as references.

Table~\ref{tab:mk_diagnostics} reports these two diagnostics for DFGO (MAE) and CDFGO-Combined on MK. Under the Gaussian reference (0.27\% exceedance at $3\sigma$ and 68.27\% coverage at $1\sigma$), DFGO (MAE) is far from the nominal values on both axes, whereas CDFGO-Combined moves much closer. This contrast shows that the reported uncertainty of CDFGO-Combined is better aligned with the realized axis-wise errors.

\begin{table*}[!htb]
    \centering
    \caption{Axis-wise uncertainty credibility diagnostics on the MK data.}
    \label{tab:mk_diagnostics}
    \begin{tabular}{@{}llcc@{}}
        \toprule
        Model & Axis & $\mathbb{P}(|e| > 3\hat{\sigma})$ & $\mathbb{P}(|e| \le 1\hat{\sigma})$ \\
        \midrule
        DFGO (MAE) & East  & 44.72\% & 21.59\% \\
        DFGO (MAE) & North & 53.88\% & 19.67\% \\
        CDFGO-Combined & East  & 10.03\% & 56.99\% \\
        CDFGO-Combined & North & 10.38\% & 53.88\% \\
        \bottomrule
    \end{tabular}
\end{table*}

\begin{figure*}[!htb]
    \centering
    \includegraphicssafe[width=0.95\textwidth]
    {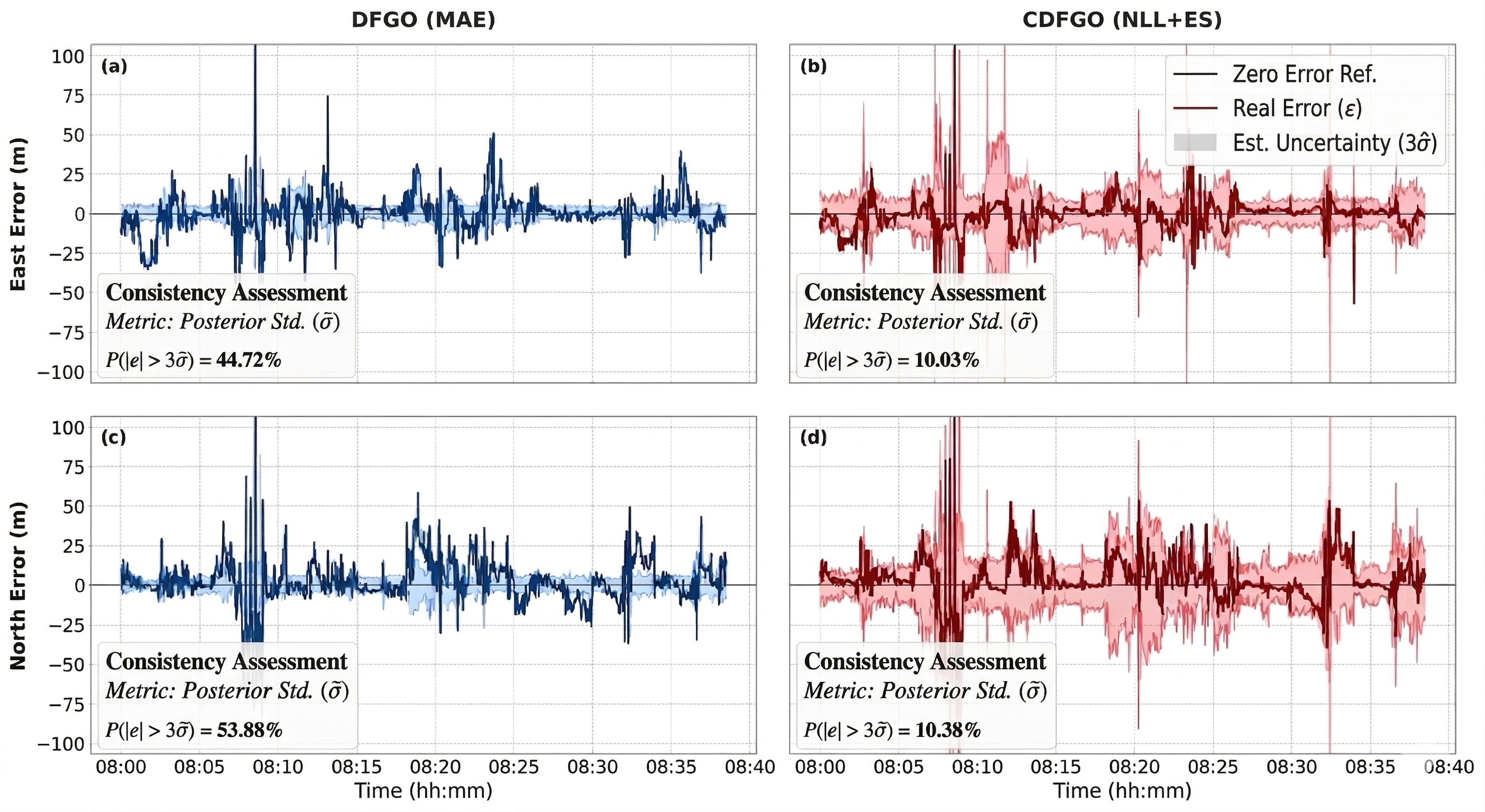}
    \caption{Full-data credibility diagnostics on MK. Panels (a) and (b) show the East-axis error for DFGO (MAE) and CDFGO-Combined, respectively, while panels (c) and (d) show the North-axis error. The colored solid curves denote realized errors, the black horizontal line denotes the zero-error reference, and the shaded regions denote the corresponding $\pm 3\hat{\sigma}$ envelopes from posterior covariance.}
    \label{fig:mk_consistency}
\end{figure*}

Figure~\ref{fig:mk_consistency} shows the same comparison over the full MK test run. Panels (a) and (c) show frequent East- and North-axis excursions beyond the reported envelope for DFGO (MAE). Panels (b) and (d) show that CDFGO-Combined widens the envelope during difficult intervals and reduces these violations. The run-level pattern is therefore consistent with the statistics in Table~\ref{tab:mk_diagnostics}.

Table~\ref{tab:mk_diagnostics} and Figure~\ref{fig:mk_consistency} show that CDFGO-Combined improves the match between reported EN uncertainty and realized EN error on MK, but they do not yet show how that mismatch appears in local geometry. Therefore, Section~\ref{subsec:mk_map_case} performs a case study to investigate the performance regarding local geometry

\subsection{Local Map-Domain Case Study on the MK Data}
\label{subsec:mk_map_case}

\begin{figure*}[!htb]
    \centering
    \begin{subfigure}[t]{0.48\textwidth}
        \centering
        \includegraphicssafe[width=\linewidth]{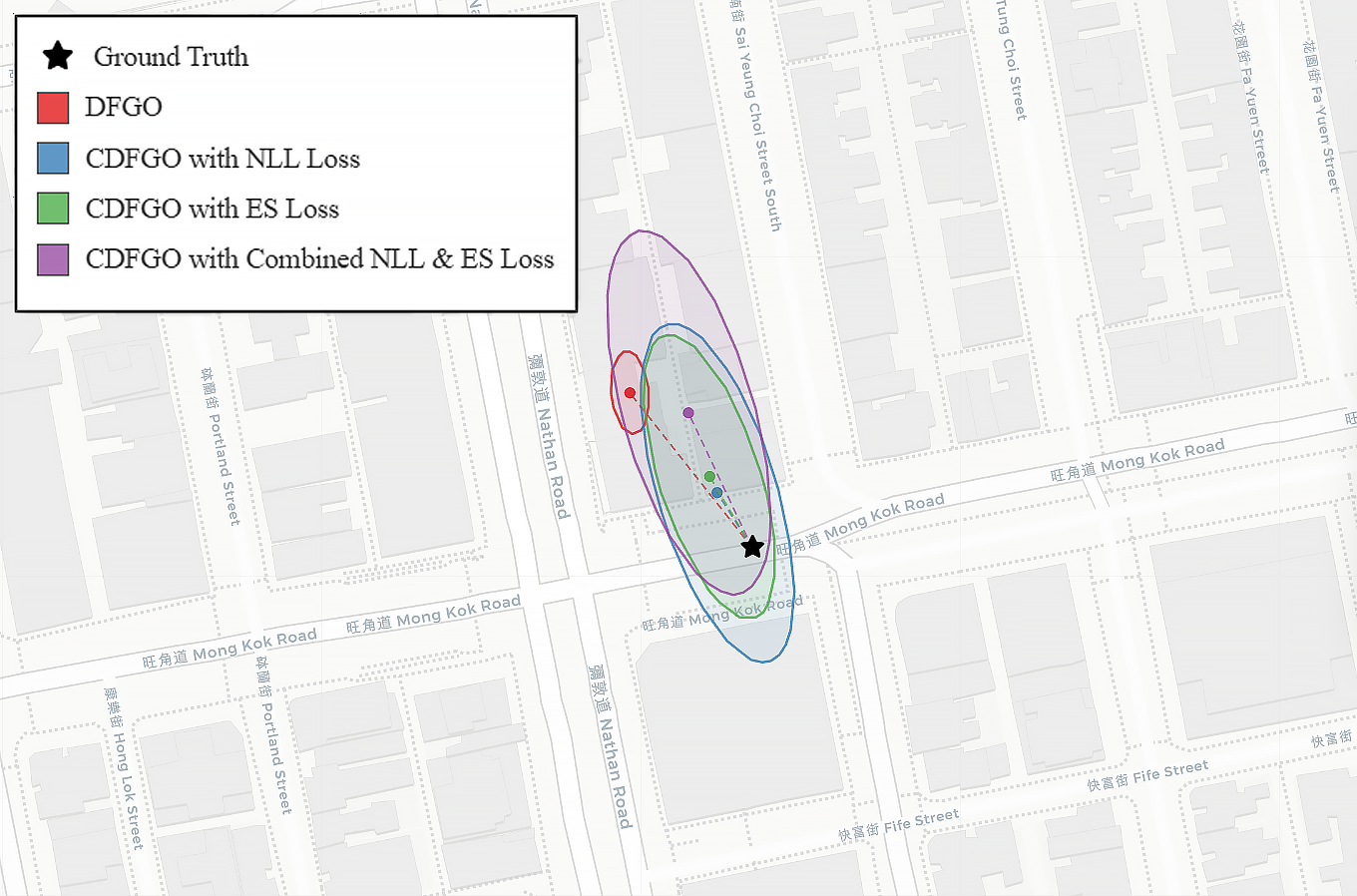}
        \caption{Selected epoch (epoch~1904).}
        \label{fig:mk_map_case_a}
    \end{subfigure}
    \hfill
    \begin{subfigure}[t]{0.48\textwidth}
        \centering
        \includegraphicssafe[width=\linewidth]{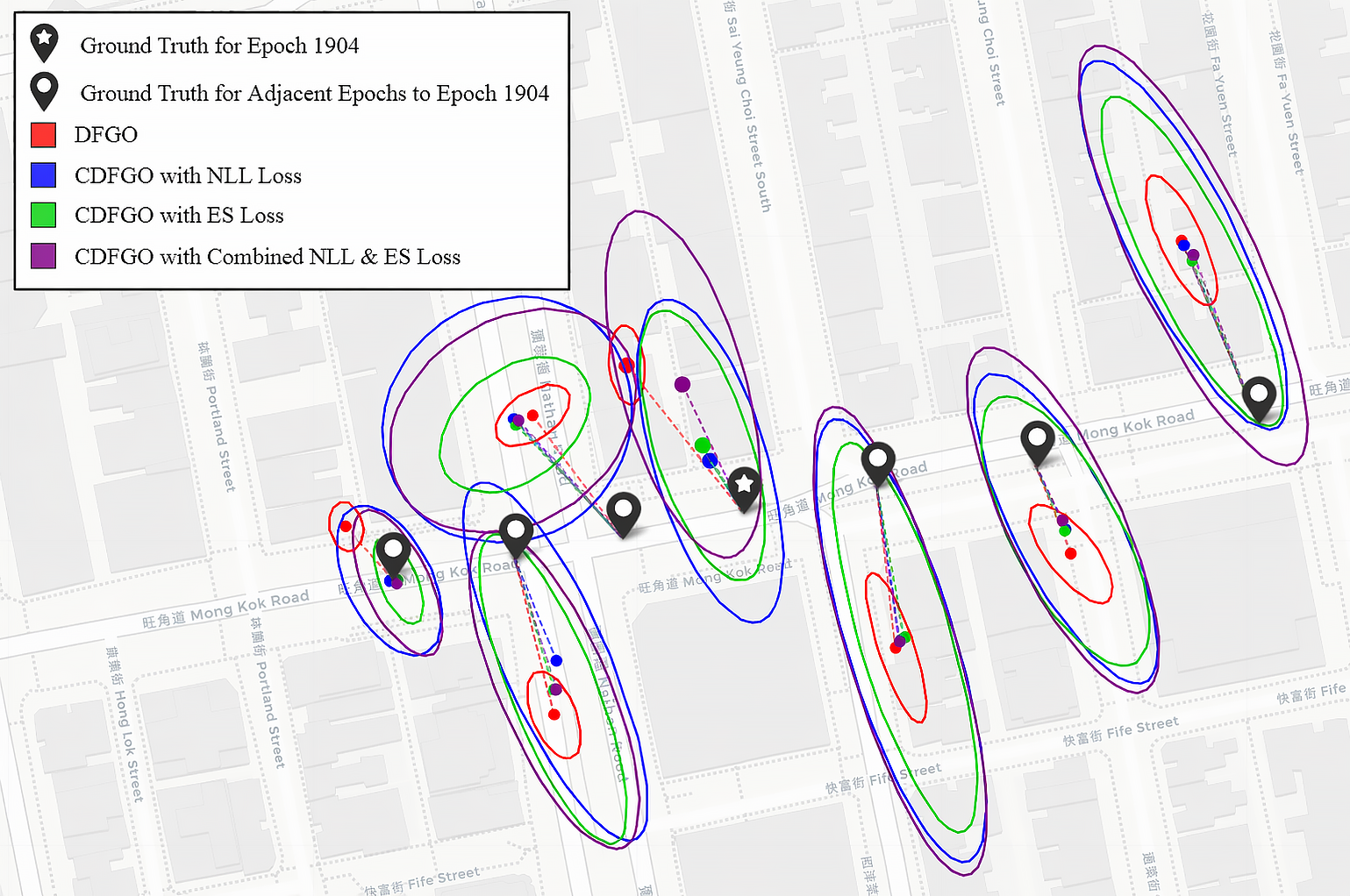}
        \caption{Adjacent epochs on the same road segment.}
        \label{fig:mk_map_case_b}
    \end{subfigure}
    \caption{Map-domain comparison for the selected MK epoch. In both panels, the colored markers and ellipses denote the horizontal estimates and the corresponding EN posterior covariance. Black stars mark the ground-truth positions.}
    \label{fig:mk_map_case}
\end{figure*}

We examine a selected MK epoch on a local map to show how the mismatch between reported EN uncertainty and realized EN error appears on the horizontal plane. Figure~\ref{fig:mk_map_case}(a) plots the horizontal estimates and EN uncertainty ellipses at epoch~1904. Figure~\ref{fig:mk_map_case}(b) extends the same comparison to several adjacent epochs on the same road segment.

In Fig.~\ref{fig:mk_map_case}(a), DFGO (MAE) gives a tight ellipse whose center is visibly shifted from the ground-truth position. By contrast, the three CDFGO variants produce larger and more anisotropic ellipses whose centers remain closer to the ground truth. The joint change in ellipse size, orientation, and aspect ratio suggests that the reported EN covariance is reshaped rather than merely inflated. Figure~\ref{fig:mk_map_case}(b) shows that the same pattern remains visible at adjacent epochs. DFGO (MAE) keeps producing tight ellipses with a persistent local offset, whereas the CDFGO ellipses are broader and change more gradually in size and orientation along the local trajectory. This map-domain behavior agrees with the axis-wise diagnostics in Table~\ref{tab:mk_diagnostics}.

\subsection{Satellite-Level Mechanism Analysis on the MK Data}
\label{subsec:mk_satellite_case}

To explain this reshaping of the reported EN covariance, we examine satellite-level weights at the same MK epoch. We compare the learned weights with two satellite-level quantities. The first is the normalized weight in Eq.~\eqref{eq:norm_weight}
\begin{equation}
\bar{w}_i = \frac{w_i}{\sum_{j \in \mathcal{I}_t} w_j},
\qquad
\sum_{i \in \mathcal{I}_t} \bar{w}_i = 1,
\label{eq:norm_weight}
\end{equation}
where $\mathcal{I}_t$ is the set of visible pseudorange factors at the selected epoch. The second quantity is a ground-truth-referenced single-differenced pseudorange error:
\begin{equation}
\tilde{\epsilon}_i = \rho_i^{\mathrm{obs}} - \left(\|\vect{p}^{gt} - \vect{s}_i\|_2 + \vect{u}_i^{\top}\vect{b}_t + \Delta\rho_i^{\mathrm{corr}}\right),
\qquad
e_i^{\mathrm{SD}} = \tilde{\epsilon}_i - \tilde{\epsilon}_{i^\star},
\label{eq:sd_error}
\end{equation}
where $\vect{p}^{gt}$ is the ground-truth receiver position at the selected epoch, $\vect{s}_i$ is the satellite position, $\vect{u}_i$ and $\vect{b}_t$ are the constellation-selection vector and the receiver clock-bias vector defined in Section~\ref{subsec:problem_setup}, $\Delta\rho_i^{\mathrm{corr}}$ collects the standard corrections, and $i^\star$ is the reference satellite in the same constellation, chosen as the one with the highest $C/N_0$. Because $\vect{u}_i=\vect{u}_{i^\star}$ within one constellation, the clock term $\vect{u}_i^{\top}\vect{b}_t$ is canceled by the differencing step. Therefore, the preliminary WLS residual $r_{t,i}$ used as a WGN feature in Section~\ref{subsec:wgn_new} and the single-differenced error $e_i^{\mathrm{SD}}$ in Eq.~\eqref{eq:sd_error} play different roles: the WLS residual reports the local fit of the preliminary solution, while the single-differenced error references the ground-truth position and measures relative contamination with respect to the reference satellite.

\begin{figure*}[!htb]
    \centering
    \includegraphicssafe[width=0.95\textwidth]{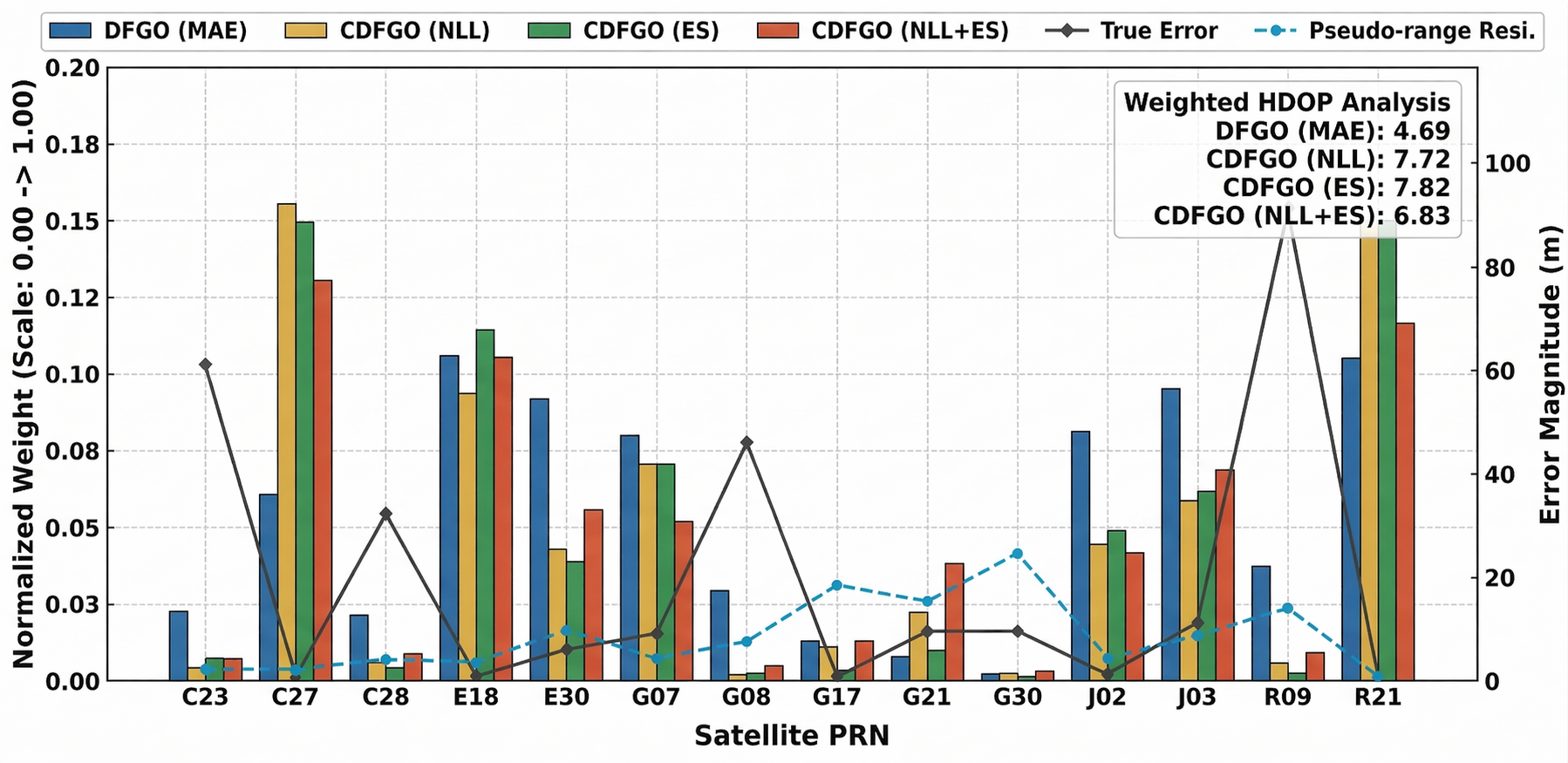}
    \caption{Per-satellite diagnosis for the selected epoch (epoch~1904) on the MK data. Bars show the normalized weights $\bar{w}_i$. The black curve is the ground-truth-referenced single-differenced error magnitude $|e_i^{\mathrm{SD}}|$, and the blue dashed curve is the WLS residual magnitude $|r_{t,i}|$. The reference satellite per constellation is omitted because its single-differenced error is zero by construction. The inset reports the weighted HDOP after reweighting.}
    \label{fig:mk_satellite_diag}
\end{figure*}

Figure~\ref{fig:mk_satellite_diag} compares the normalized weights, the ground-truth-referenced single-differenced error magnitude $|e_i^{\mathrm{SD}}|$, and the WLS residual magnitude $|r_{t,i}|$ at the selected epoch. The reference satellite of each constellation is not shown because its single-differenced error is zero by construction. Several satellites illustrate why residual size alone is not sufficient to identify contaminated measurements. For example, C23, G08, and R09 have small or moderate WLS residuals, yet their single-differenced errors remain large. DFGO (MAE) still assigns non-negligible weight to these satellites, whereas the CDFGO variants, especially CDFGO-Combined, suppress them more aggressively. Compared with DFGO (MAE), the pattern suggests that the credibility objective lowers the weight of measurements whose residual fit still looks acceptable even when their ground-truth-referenced contamination remains high.

The inset in Fig.~\ref{fig:mk_satellite_diag} shows the cost of this choice at the selected epoch. The weighted horizontal dilution of precision (HDOP) rises from 4.69 for DFGO (MAE) to 7.72, 7.82, and 6.83 for CDFGO-NLL, CDFGO-ES, and CDFGO-Combined, respectively. Relative to DFGO (MAE), the credibility-driven variants therefore accept weaker nominal geometry after contaminated satellites are suppressed. However, Table~\ref{tab:main_results} and Table~\ref{tab:mk_diagnostics} show that, on MK, this trade still yields lower mean and 95th-percentile errors together with lower NLL and ES than DFGO (MAE). On the same scene, the larger covariance reported by CDFGO-Combined is therefore better aligned with the realized positioning risk than the smaller covariance from DFGO (MAE).

\begin{figure*}[!htb]
    \centering
    \includegraphicssafe[width=0.95\textwidth]{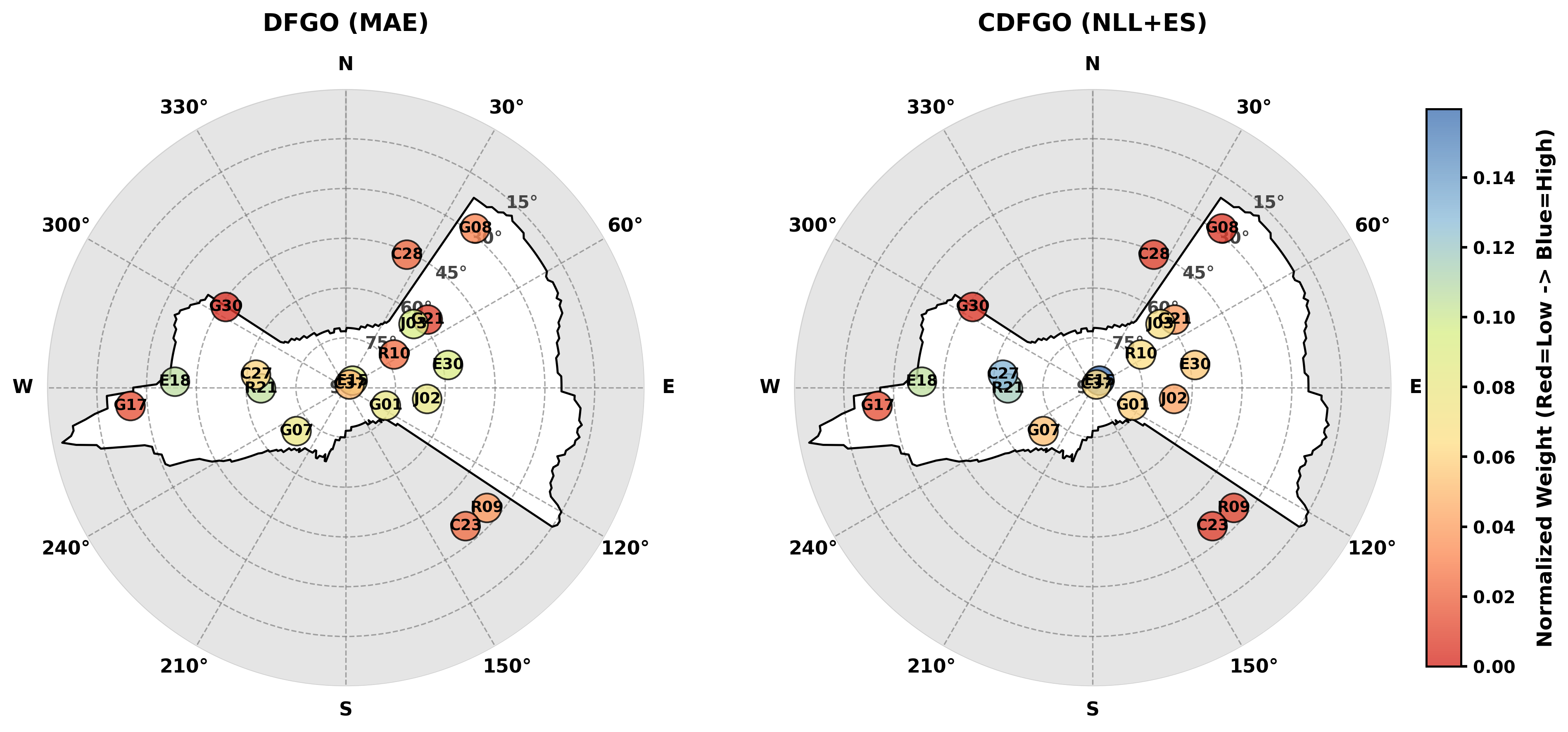}
    \caption{Skyplot-level trust diagnosis for the selected epoch (epoch~1904) on the MK data. The left panel shows DFGO (MAE), and the right panel shows CDFGO-Combined. Marker color encodes the normalized weight $\bar{w}_i$, with warmer colors indicating lower trust. Comparing the two panels shows how the credibility objective redistributes weight across azimuth.}
    \label{fig:mk_skyplot_diag}
\end{figure*}

However, weight suppression alone does not explain the full MK behavior. The solution also depends on which satellites remain trusted after reweighting. Therefore, we plot the skyplot of satellites in this epoch in Figure~\ref{fig:mk_skyplot_diag} with marker color encoding the normalized weight. The satellites are unevenly distributed across azimuth, so horizontal support is direction-dependent. Compared with DFGO (MAE), CDFGO-Combined assigns less weight to suspicious satellites such as C23 and R09, while keeping relatively higher weight on satellites such as C27, R21, and E18 that support the sparser side of the skyplot. In this epoch, the reweighting does not simply discard satellites; it also keeps satellites that matter most for horizontal observability in the weak direction. This preservation partly offsets the geometry loss caused by rejecting contaminated factors.

At this epoch, CDFGO-Combined lowers weight on satellites whose WLS residual hides their true contamination (Figure~\ref{fig:mk_satellite_diag}), while keeping higher trust on satellites that fill the weak azimuth sector (Figure~\ref{fig:mk_skyplot_diag}). These two adjustments together explain the larger map-domain covariance in Figure~\ref{fig:mk_map_case} and the lower exceedance and higher in-envelope coverage reported for CDFGO-Combined in Section~\ref{subsec:credibility_diagnostics}.

\FloatBarrier
\section{Discussion and Limitations}
\label{sec:discussion_new}

The experiment results in Section~\ref{sec:experiments_new} indicate that credibility training mainly changes how the solver's covariance can be used downstream. In the present setting, covariance is no longer a passive byproduct of the solver; it becomes an explicit training target. This matters because downstream fusion and integrity modules act on covariance, not on horizontal error alone. In multi-sensor fusion, covariance controls the update gain and therefore how strongly GNSS pulls the fused estimate \citep{wen2021fgoins}. In integrity-oriented processing, uncertainty quality is itself part of the decision logic \citep{xia2024integrity}. A GNSS backend that reports more credible covariance is therefore a stronger candidate for downstream fusion and integrity integration, even though those downstream systems are not evaluated in this work.

The credibility-oriented objectives also provide a direct training-time diagnostic. Because the uncertainty behavior is tied to an explicit proper score, a model update can be judged not only by aggregate positioning error but also by whether the reported covariance remains statistically consistent with realized errors, which is exactly the role of proper scoring rules such as NLL \citep{gneiting2007}. The exceedance and coverage diagnostics in Section~\ref{subsec:credibility_diagnostics} serve the same purpose for credibility checking \citep{li2012credibility}. They do not provide formal certification, but they do reveal whether the uncertainty behavior is moving toward statistical consistency. This framing is closer to integrity-oriented evaluation than to optimizing geometry-oriented losses alone.

At the same time, this study has clear limits. The loss balancing still uses fixed coefficients $(\alpha,\beta)$, although the calibration-sharpness trade-off can vary with scene difficulty and mission profile. The posterior covariance is still extracted from a local Laplace approximation, which is efficient but may be insufficient when the posterior is strongly multi-modal. These two points suggest natural extensions toward adaptive loss balancing and richer uncertainty families.

The experimental scope is also deliberately narrow. The reported results use pseudorange-only factors. This isolates the mechanism of credibility training, but it does not show how the same credibility-driven training objective behaves after coupling with Doppler, carrier phase, or inertial constraints. Extending the framework to tightly coupled GNSS/INS or GNSS/vision graphs is a necessary next step for studying downstream fusion and integrity behavior \citep{wen2021fgoins,xia2024integrity}. The mechanism interpretation in Section~\ref{subsec:mk_satellite_case} is also built around a representative MK epoch. This selected epoch is useful for explanation, but it does not establish the same causal pattern across all epochs and scenes. Larger-scale satellite-level statistics are still needed, and developing such statistics is a clear direction for future work.

The present evaluation is further limited by geography, hardware, and reference quality. UrbanNav covers several urban scenes \citep{hsu2023urbannav}, but its reference trajectories are tied to a single hardware setup and to high-grade RTK/INS post-processing that may still retain some uncertainty under severe blockage. Because the credibility losses depend on reference quality, label errors can transfer into the learned weighting behavior. End-to-end differentiable optimization with Monte Carlo ES estimation is also more expensive than hand-crafted weighting rules, and gradient stability still depends on solver convergence. Broader cross-domain experiments, more diverse hardware platforms, and more efficient training strategies are therefore needed before deployment-oriented use can be claimed.

\section{Conclusion}
\label{sec:conclusion_new}

In this work, we introduced CDFGO, a differentiable GNSS factor graph trained with proper-scoring-rule losses on its East--North predictive distribution. Experiments on the UrbanNav dataset across medium, deep, and harsh urban scenes confirmed that credibility supervision produces a more reliable posterior covariance and improves horizontal accuracy. In contrast, the DFGO (MAE) baseline with a position-only loss returns miscalibrated covariance even when its position estimate is accurate. This work offers GNSS practitioners a way to train covariance as a primary estimator output, enabling direct use in downstream fusion and integrity monitoring.

\bibliographystyle{plainnat}
\bibliography{references_v2}

\begin{thebibliography}{51}
\providecommand{\natexlab}[1]{#1}
\providecommand{\url}[1]{\texttt{#1}}
\expandafter\ifx\csname urlstyle\endcsname\relax
  \providecommand{\doi}[1]{doi: #1}\else
  \providecommand{\doi}{doi: \begingroup \urlstyle{rm}\Url}\fi

\bibitem[Amos and Kolter(2017)]{amos2017optnet}
Brandon Amos and J.~Zico Kolter.
\newblock Optnet: Differentiable optimization as a layer in neural networks.
\newblock In \emph{Proceedings of the 34th International Conference on Machine
  Learning (ICML)}, volume~70 of \emph{Proceedings of Machine Learning
  Research}, pages 136--145. PMLR, 2017.
\newblock \doi{10.48550/arXiv.1703.00443}.

\bibitem[Chen et~al.(2023)Chen, Sun, Fu, Cheng, and
  Chiang]{chen2023pseudorangecorrection}
Wu~Chen, Rui Sun, Linxia Fu, Qi~Cheng, and Kai~Wei Chiang.
\newblock Resilient pseudorange error prediction and correction for gnss
  positioning in urban areas.
\newblock \emph{IEEE Internet of Things Journal}, 10\penalty0 (11):\penalty0
  9979--9988, 2023.
\newblock \doi{10.1109/JIOT.2023.3235483}.

\bibitem[Dellaert and Kaess(2017)]{dellaert2017factorgraphs}
Frank Dellaert and Michael Kaess.
\newblock Factor graphs for robot perception.
\newblock \emph{Foundations and Trends in Robotics}, 6\penalty0
  (1--2):\penalty0 1--139, 2017.
\newblock \doi{10.1561/2300000043}.

\bibitem[Gal and Ghahramani(2016)]{gal2016dropout}
Yarin Gal and Zoubin Ghahramani.
\newblock Dropout as a bayesian approximation: Representing model uncertainty
  in deep learning.
\newblock In \emph{Proceedings of the 33rd International Conference on Machine
  Learning (ICML)}, volume~48 of \emph{Proceedings of Machine Learning
  Research}, pages 1050--1059. PMLR, 2016.
\newblock \doi{10.48550/arXiv.1506.02142}.

\bibitem[Gallon et~al.(2022)Gallon, Joerger, and Pervan]{gallon2022orbitclock}
Elisa Gallon, Mathieu Joerger, and Boris Pervan.
\newblock Robust modeling of gnss orbit and clock error dynamics.
\newblock \emph{NAVIGATION: Journal of the Institute of Navigation},
  69\penalty0 (4):\penalty0 navi.539, 2022.
\newblock \doi{10.33012/navi.539}.

\bibitem[Garcia~Crespillo et~al.(2020)Garcia~Crespillo, Andreetti, and
  Grosch]{crespillo2020mestimators}
Omar Garcia~Crespillo, Alice Andreetti, and Anja Grosch.
\newblock Design and evaluation of robust m-estimators for gnss positioning in
  urban environments.
\newblock In \emph{Proceedings of the 2020 International Technical Meeting of
  The Institute of Navigation}, pages 750--762. Institute of Navigation, 2020.
\newblock \doi{10.33012/2020.17211}.

\bibitem[Gneiting and Raftery(2007)]{gneiting2007}
Tilmann Gneiting and Adrian~E. Raftery.
\newblock Strictly proper scoring rules, prediction, and estimation.
\newblock \emph{Journal of the American Statistical Association}, 102\penalty0
  (477):\penalty0 359--378, 2007.
\newblock \doi{10.1198/016214506000001437}.

\bibitem[Gneiting et~al.(2008)Gneiting, Stanberry, Grimit, Held, and
  Johnson]{gneiting2008}
Tilmann Gneiting, Larissa~I. Stanberry, Eric~P. Grimit, Leonhard Held, and
  Nicholas~A. Johnson.
\newblock Assessing probabilistic forecasts of multivariate quantities, with an
  application to ensemble predictions of surface winds.
\newblock \emph{TEST}, 17\penalty0 (2):\penalty0 211--235, 2008.
\newblock \doi{10.1007/s11749-008-0114-x}.

\bibitem[Hartinger and Brunner(1999)]{hartinger1999sigma}
H.~Hartinger and F.~K. Brunner.
\newblock Variances of {GPS} phase observations: The {SIGMA}-$\varepsilon$
  model.
\newblock \emph{GPS Solutions}, 2\penalty0 (4):\penalty0 35--43, 1999.
\newblock \doi{10.1007/pl00012765}.

\bibitem[Hsu et~al.(2021)Hsu, Kubo, Wen, Chen, Liu, Suzuki, and
  Meguro]{hsu2021urbannav}
Li-Ta Hsu, Nobuaki Kubo, Weisong Wen, Wu~Chen, Zhizhao Liu, Taro Suzuki, and
  Junichi Meguro.
\newblock Urbannav: An open-sourced multisensory dataset for benchmarking
  positioning algorithms designed for urban areas.
\newblock In \emph{Proceedings of the 34th International Technical Meeting of
  the Satellite Division of The Institute of Navigation (ION GNSS+ 2021)},
  pages 226--256. Institute of Navigation, 2021.
\newblock \doi{10.33012/2021.17895}.

\bibitem[Hsu et~al.(2023)Hsu, Huang, Ng, Zhang, Zhong, Bai, and
  Wen]{hsu2023urbannav}
Li-Ta Hsu, Feng Huang, Hoi-Fung Ng, Guohao Zhang, Yihan Zhong, Xiwei Bai, and
  Weisong Wen.
\newblock Hong kong urbannav: An open-source multisensory dataset for
  benchmarking urban navigation algorithms.
\newblock \emph{NAVIGATION: Journal of the Institute of Navigation},
  70\penalty0 (4):\penalty0 navi.602, 2023.
\newblock \doi{10.33012/navi.602}.

\bibitem[Hu et~al.(2025)Hu, Xu, Zhong, and Wen]{hu2025tdlgnss}
Runzhi Hu, Penghui Xu, Yihan Zhong, and Weisong Wen.
\newblock pyrtklib: An open-source package for tightly coupled deep learning
  and gnss integration for positioning in urban canyons.
\newblock \emph{IEEE Transactions on Intelligent Transportation Systems},
  26\penalty0 (7):\penalty0 10652--10662, 2025.
\newblock \doi{10.1109/TITS.2025.3552691}.

\bibitem[Joerger and Pervan(2020)]{joerger2020araimmotion}
Mathieu Joerger and Boris Pervan.
\newblock Multi-constellation araim exploiting satellite motion.
\newblock \emph{NAVIGATION: Journal of the Institute of Navigation},
  67\penalty0 (2):\penalty0 235--253, 2020.
\newblock \doi{10.1002/navi.334}.

\bibitem[Kaess et~al.(2012)Kaess, Johannsson, Roberts, Ila, Leonard, and
  Dellaert]{kaess2012isam2}
Michael Kaess, Hordur Johannsson, Richard Roberts, Viorela Ila, John~J.
  Leonard, and Frank Dellaert.
\newblock {iSAM2}: Incremental smoothing and mapping using the bayes tree.
\newblock \emph{The International Journal of Robotics Research}, 31\penalty0
  (2):\penalty0 217--236, 2012.
\newblock \doi{10.1177/0278364911430419}.

\bibitem[Kanhere et~al.(2022)Kanhere, Gupta, Shetty, and
  Gao]{kanhere2022nncorrections}
Ashwin~V. Kanhere, Shubh Gupta, Akshay Shetty, and Grace Gao.
\newblock Improving gnss positioning using neural-network-based corrections.
\newblock \emph{NAVIGATION: Journal of the Institute of Navigation},
  69\penalty0 (4):\penalty0 navi.548, 2022.
\newblock \doi{10.33012/navi.548}.

\bibitem[Kaplan and Hegarty(2017)]{kaplan2017gps}
Elliott~D. Kaplan and Christopher Hegarty.
\newblock \emph{Understanding GPS/GNSS: Principles and Applications}.
\newblock Artech House, Norwood, MA, 3 edition, 2017.
\newblock ISBN 9781630810580.

\bibitem[Kingma and Ba(2015)]{kingma2014adam}
Diederik~P. Kingma and Jimmy Ba.
\newblock Adam: A method for stochastic optimization.
\newblock In \emph{Proceedings of the 3rd International Conference on Learning
  Representations (ICLR)}, 2015.
\newblock \doi{10.48550/arXiv.1412.6980}.

\bibitem[Kingma and Welling(2014)]{kingma2014vae}
Diederik~P. Kingma and Max Welling.
\newblock Auto-encoding variational bayes.
\newblock In \emph{International Conference on Learning Representations
  (ICLR)}, 2014.
\newblock \doi{10.48550/arXiv.1312.6114}.
\newblock URL \url{https://arxiv.org/abs/1312.6114}.

\bibitem[Knowles and Gao(2023)]{knowles2023edmfde}
Derek Knowles and Grace Gao.
\newblock Euclidean distance matrix-based rapid fault detection and exclusion.
\newblock \emph{NAVIGATION: Journal of the Institute of Navigation},
  70\penalty0 (1):\penalty0 navi.555, 2023.
\newblock \doi{10.33012/navi.555}.

\bibitem[Kschischang et~al.(2001)Kschischang, Frey, and
  Loeliger]{kschischang2001factor}
Frank~R. Kschischang, Brendan~J. Frey, and Hans-Andrea Loeliger.
\newblock Factor graphs and the sum-product algorithm.
\newblock \emph{IEEE Transactions on Information Theory}, 47\penalty0
  (2):\penalty0 498--519, 2001.
\newblock \doi{10.1109/18.910572}.

\bibitem[Levi et~al.(2022)Levi, Gispan, Giladi, and
  Fetaya]{levi2022regressioncalibration}
Dan Levi, Liran Gispan, Niv Giladi, and Ethan Fetaya.
\newblock Evaluating and calibrating uncertainty prediction in regression
  tasks.
\newblock \emph{Sensors}, 22\penalty0 (15):\penalty0 5540, 2022.
\newblock \doi{10.3390/s22155540}.

\bibitem[Li et~al.(2023)Li, Elhajj, Feng, and Ochieng]{li2023nlosweighting}
Lintong Li, Mireille Elhajj, Yuxiang Feng, and Washington~Yotto Ochieng.
\newblock Machine learning based gnss signal classification and weighting
  scheme design in the built environment: a comparative experiment.
\newblock \emph{Satellite Navigation}, 4:\penalty0 12, 2023.
\newblock \doi{10.1186/s43020-023-00101-w}.

\bibitem[Li et~al.(2012)Li, Zhao, and Li]{li2012credibility}
X.~Rong Li, Zhanlue Zhao, and Xiao-Bai Li.
\newblock Evaluation of estimation algorithms: Credibility tests.
\newblock \emph{IEEE Transactions on Systems, Man, and Cybernetics - Part A:
  Systems and Humans}, 42\penalty0 (1):\penalty0 147--163, 2012.
\newblock \doi{10.1109/TSMCA.2011.2158095}.

\bibitem[Li et~al.(2026)Li, Yan, and Hsu]{li2026lah}
Zhengdao Li, Penggao Yan, and Li-Ta Hsu.
\newblock Logistic-aided {Huber} {M}-estimator for robust {GNSS} positioning.
\newblock arXiv preprint arXiv:2603.19640, 2026.

\bibitem[Medina et~al.(2019)Medina, Li, Vila-Valls, and
  Closas]{medina2019robuststats}
Daniel Medina, Haoqing Li, Jordi Vila-Valls, and Pau Closas.
\newblock Robust statistics for gnss positioning under harsh conditions: A
  useful tool?
\newblock \emph{Sensors}, 19\penalty0 (24):\penalty0 5402, 2019.
\newblock \doi{10.3390/s19245402}.

\bibitem[Mohanty and Gao(2023)]{mohanty2023gcnncorrections}
Adyasha Mohanty and Grace Gao.
\newblock Learning {GNSS} positioning corrections for smartphones using graph
  convolution neural networks.
\newblock \emph{NAVIGATION: Journal of the Institute of Navigation},
  70\penalty0 (4):\penalty0 navi.622, 2023.
\newblock \doi{10.33012/navi.622}.

\bibitem[Mohanty and Gao(2024)]{mohanty2024gnnkf}
Adyasha Mohanty and Grace Gao.
\newblock Tightly coupled graph neural network and kalman filter for smartphone
  positioning.
\newblock \emph{NAVIGATION: Journal of the Institute of Navigation},
  71\penalty0 (4):\penalty0 navi.670, 2024.
\newblock \doi{10.33012/navi.670}.

\bibitem[Nagai et~al.(2024)Nagai, Spenko, Henderson, and
  Pervan]{nagai2024faultfreeintegrity}
Kana Nagai, Matthew Spenko, Ron Henderson, and Boris Pervan.
\newblock Fault-free integrity of urban driverless vehicle navigation with
  multi-sensor integration: A case study in downtown chicago.
\newblock \emph{NAVIGATION: Journal of the Institute of Navigation},
  71\penalty0 (1):\penalty0 navi.631, 2024.
\newblock \doi{10.33012/navi.631}.

\bibitem[Ng et~al.(2021)Ng, Zhang, Luo, and Hsu]{ng2021smartphone3dma}
Hoi-Fung Ng, Guohao Zhang, Yiran Luo, and Li-Ta Hsu.
\newblock Urban positioning: 3d mapping-aided gnss using dual-frequency
  pseudorange measurements from smartphones.
\newblock \emph{NAVIGATION: Journal of the Institute of Navigation},
  68\penalty0 (4):\penalty0 727--749, 2021.
\newblock \doi{10.1002/navi.448}.

\bibitem[Pineda et~al.(2022)Pineda, Fan, Monge, Venkataraman, Sodhi, Chen,
  Ortiz, DeTone, Wang, Anderson, Dong, Amos, and Mukadam]{theseus2022}
Luis Pineda, Taosha Fan, Maurizio Monge, Shobha Venkataraman, Paloma Sodhi,
  Ricky T.~Q. Chen, Joseph Ortiz, Daniel DeTone, Austin Wang, Stuart Anderson,
  Jing Dong, Brandon Amos, and Mustafa Mukadam.
\newblock Theseus: A library for differentiable nonlinear optimization.
\newblock In \emph{Advances in Neural Information Processing Systems
  ({NeurIPS})}, volume~35, 2022.
\newblock \doi{10.48550/arXiv.2207.09442}.
\newblock URL
  \url{https://papers.nips.cc/paper_files/paper/2022/hash/185969291540b3cd86e70c51e8af5d08-Abstract-Conference.html}.

\bibitem[Realini and Reguzzoni(2013)]{realini2013gogps}
Eugenio Realini and Mirko Reguzzoni.
\newblock go{GPS}: Open source software for enhancing the accuracy of low-cost
  receivers by single-frequency relative kinematic positioning.
\newblock \emph{Measurement Science and Technology}, 24\penalty0 (11):\penalty0
  115010, 2013.
\newblock \doi{10.1088/0957-0233/24/11/115010}.

\bibitem[Rezende et~al.(2014)Rezende, Mohamed, and
  Wierstra]{rezende2014stochastic}
Danilo~Jimenez Rezende, Shakir Mohamed, and Daan Wierstra.
\newblock Stochastic backpropagation and approximate inference in deep
  generative models.
\newblock In \emph{Proceedings of the 31st International Conference on Machine
  Learning}, volume~32 of \emph{Proceedings of Machine Learning Research},
  pages 1278--1286. PMLR, 2014.
\newblock \doi{10.48550/arXiv.1401.4082}.
\newblock URL \url{https://proceedings.mlr.press/v32/rezende14.html}.

\bibitem[Scheuerer and Hamill(2015)]{scheuerer2015}
Michael Scheuerer and Thomas~M. Hamill.
\newblock Variogram-based proper scoring rules for probabilistic forecasts of
  multivariate quantities.
\newblock \emph{Monthly Weather Review}, 143\penalty0 (4):\penalty0 1321--1334,
  2015.
\newblock \doi{10.1175/MWR-D-14-00269.1}.

\bibitem[Smolyakov et~al.(2020)Smolyakov, Rezaee, and
  Langley]{smolyakov2020multipathprediction}
Ivan Smolyakov, Mohammad Rezaee, and Richard~B. Langley.
\newblock Resilient multipath prediction and detection architecture for
  low-cost navigation in challenging urban areas.
\newblock \emph{NAVIGATION: Journal of the Institute of Navigation},
  67\penalty0 (2):\penalty0 397--409, 2020.
\newblock \doi{10.1002/navi.362}.

\bibitem[Taylor and Gross(2024)]{taylor2024factorgraphs}
Clark Taylor and Jason Gross.
\newblock Factor graphs for navigation applications: A tutorial.
\newblock \emph{NAVIGATION: Journal of the Institute of Navigation},
  71\penalty0 (3):\penalty0 navi.653, 2024.
\newblock \doi{10.33012/navi.653}.

\bibitem[Teunissen and Montenbruck(2017)]{teunissen2017handbook}
Peter~J.G. Teunissen and Oliver Montenbruck, editors.
\newblock \emph{Springer Handbook of Global Navigation Satellite Systems}.
\newblock Springer, Cham, Switzerland, 2017.
\newblock \doi{10.1007/978-3-319-42928-1}.

\bibitem[Vaswani et~al.(2017)Vaswani, Shazeer, Parmar, Uszkoreit, Jones, Gomez,
  Kaiser, and Polosukhin]{vaswani2017}
Ashish Vaswani, Noam Shazeer, Niki Parmar, Jakob Uszkoreit, Llion Jones,
  Aidan~N. Gomez, Lukasz Kaiser, and Illia Polosukhin.
\newblock Attention is all you need.
\newblock In \emph{Advances in Neural Information Processing Systems
  ({NeurIPS})}, volume~30, 2017.
\newblock \doi{10.48550/arXiv.1706.03762}.
\newblock URL
  \url{https://proceedings.neurips.cc/paper_files/paper/2017/file/3f5ee243547dee91fbd053c1c4a845aa-Abstract.html}.

\bibitem[Wang et~al.(2013)Wang, Groves, and Ziebart]{wang2013shadowmatching}
Lei Wang, Paul~D. Groves, and Marek~K. Ziebart.
\newblock Gnss shadow matching: Improving urban positioning accuracy using a 3d
  city model with optimized visibility scoring scheme.
\newblock \emph{NAVIGATION: Journal of the Institute of Navigation},
  60\penalty0 (3):\penalty0 195--207, 2013.
\newblock \doi{10.1002/navi.38}.

\bibitem[Wen and Hsu(2021)]{wen2021icra}
Weisong Wen and Li-Ta Hsu.
\newblock Towards robust gnss positioning and real-time kinematic using factor
  graph optimization.
\newblock In \emph{2021 IEEE International Conference on Robotics and
  Automation (ICRA)}, pages 5884--5890. IEEE, 2021.
\newblock \doi{10.1109/ICRA48506.2021.9562037}.

\bibitem[Wen et~al.(2020)Wen, Bai, Hsu, and Pfeifer]{wen2020gmmfusion}
Weisong Wen, Xiwei Bai, Li-Ta Hsu, and Tim Pfeifer.
\newblock Gnss/lidar integration aided by self-adaptive gaussian mixture models
  in urban scenarios: An approach robust to non-gaussian noise.
\newblock In \emph{2020 IEEE/ION Position, Location and Navigation Symposium
  (PLANS)}, pages 647--654. IEEE/ION, 2020.
\newblock \doi{10.1109/PLANS46316.2020.9110157}.

\bibitem[Wen et~al.(2021)Wen, Pfeifer, Bai, and Hsu]{wen2021fgoins}
Weisong Wen, Tim Pfeifer, Xiwei Bai, and Li-Ta Hsu.
\newblock Factor graph optimization for gnss/ins integration: A comparison with
  the extended kalman filter.
\newblock \emph{NAVIGATION: Journal of the Institute of Navigation},
  68\penalty0 (2):\penalty0 315--328, 2021.
\newblock \doi{10.1002/navi.421}.

\bibitem[Wen et~al.(2022)Wen, Zhang, and Hsu]{wen2022gnc}
Weisong Wen, Guohao Zhang, and Li-Ta Hsu.
\newblock Gnss outlier mitigation via graduated non-convexity factor graph
  optimization.
\newblock \emph{IEEE Transactions on Vehicular Technology}, 71\penalty0
  (1):\penalty0 297--310, 2022.
\newblock \doi{10.1109/TVT.2021.3130909}.

\bibitem[Xia et~al.(2024)Xia, Wen, and Hsu]{xia2024integrity}
Xiao Xia, Weisong Wen, and Li-Ta Hsu.
\newblock Integrity-constrained factor graph optimization for gnss positioning
  in urban canyons.
\newblock \emph{NAVIGATION: Journal of the Institute of Navigation},
  71\penalty0 (3):\penalty0 navi.660, 2024.
\newblock \doi{10.33012/navi.660}.

\bibitem[Xu and Hsu(2024)]{xu2024autow}
Penghui Xu and Li-Ta Hsu.
\newblock Autow: Self-supervision learning for weighting estimation in gnss
  positioning.
\newblock In \emph{Proceedings of the 37th International Technical Meeting of
  the Satellite Division of The Institute of Navigation (ION GNSS+ 2024)},
  pages 2630--2644, 2024.
\newblock \doi{10.33012/2024.19896}.

\bibitem[Xu et~al.(2023)Xu, Ng, Zhong, Zhang, Wen, Yang, and
  Hsu]{xu2023dfgo_icov}
Penghui Xu, Hoi-Fung Ng, Yihan Zhong, Guohao Zhang, Weisong Wen, Bo~Yang, and
  Li-Ta Hsu.
\newblock Differentiable factor graph optimization with intelligent covariance
  adaptation for accurate smartphone positioning.
\newblock In \emph{Proceedings of the 36th International Technical Meeting of
  the Satellite Division of The Institute of Navigation (ION GNSS+ 2023)},
  pages 2765--2773, 2023.
\newblock \doi{10.33012/2023.19297}.

\bibitem[Yan et~al.(2025{\natexlab{a}})Yan, Hu, Wen, and
  Hsu]{yan2025multifaults}
Penggao Yan, Yingjie Hu, Weisong Wen, and Li-Ta Hsu.
\newblock Multiple faults isolation for multiconstellation {GNSS} positioning
  through incremental expansion of consistent measurements.
\newblock \emph{IEEE Sensors Journal}, 25\penalty0 (4):\penalty0 6967--6980,
  February 2025{\natexlab{a}}.
\newblock \doi{10.1109/JSEN.2024.3524434}.

\bibitem[Yan et~al.(2025{\natexlab{b}})Yan, Xia, Brizzi, Wen, and
  Hsu]{yan2025subspacegmm}
Penggao Yan, Xiao Xia, Michele Brizzi, Weisong Wen, and Li-Ta Hsu.
\newblock Subspace-based adaptive gmm error modeling for fault-aware
  pseudorange-based positioning in urban canyons.
\newblock \emph{IEEE Transactions on Intelligent Vehicles}, 10\penalty0
  (5):\penalty0 3222--3237, 2025{\natexlab{b}}.
\newblock \doi{10.1109/TIV.2024.3450198}.

\bibitem[Yan et~al.(2025{\natexlab{c}})Yan, Zhan, Sun, and
  Hsu]{yan2025credibleuq}
Penggao Yan, Xingqun Zhan, Rui Sun, and Li-Ta Hsu.
\newblock Credible uncertainty quantification under noise and system model
  mismatch.
\newblock arXiv preprint arXiv:2509.03311, 2025{\natexlab{c}}.

\bibitem[Yi et~al.(2021)Yi, Lee, Kloss, Martin-Martin, and Bohg]{yi2021dfgo}
Brent Yi, Michelle~A. Lee, Alina Kloss, Roberto Martin-Martin, and Jeannette
  Bohg.
\newblock Differentiable factor graph optimization for learning smoothers.
\newblock In \emph{2021 IEEE/RSJ International Conference on Intelligent Robots
  and Systems (IROS)}, pages 1339--1345. IEEE, IEEE, 2021.
\newblock \doi{10.1109/IROS51168.2021.9636300}.

\bibitem[Zhang et~al.(2021)Zhang, Xu, Xu, and
  Hsu]{zhang2021urbangnssuncertainty}
Guohao Zhang, Penghui Xu, Haosheng Xu, and Li-Ta Hsu.
\newblock Prediction on the urban gnss measurement uncertainty based on deep
  learning networks with long short-term memory.
\newblock \emph{IEEE Sensors Journal}, 21\penalty0 (18):\penalty0 20563--20577,
  2021.
\newblock \doi{10.1109/JSEN.2021.3098006}.

\bibitem[Zheng et~al.(2024)Zheng, Zeng, Li, Wang, Xie, Liu, and
  Xie]{zheng2024visibility}
Shaolong Zheng, Kungan Zeng, Zhenni Li, Qianming Wang, Kan Xie, Ming Liu, and
  Shengli Xie.
\newblock Improving the prediction of gnss satellite visibility in urban
  canyons based on a graph transformer.
\newblock \emph{NAVIGATION: Journal of the Institute of Navigation},
  71\penalty0 (4):\penalty0 navi.676, 2024.
\newblock \doi{10.33012/navi.676}.

\end{thebibliography}

\end{document}